\documentclass[reprint,showpacs,preprintnumbers,amsmath,amssymb]{revtex4}

\usepackage{graphicx}
\usepackage{bm}
\usepackage{mathrsfs}
\usepackage{amsmath}
\usepackage{amsfonts}
\usepackage{amssymb}
\usepackage{latexsym}
\usepackage{graphicx}
\usepackage[english]{babel}
\usepackage{epsfig}
\def\be{\begin{equation}}
\def\ee{\end{equation}}

\def\pri{'}
\def\G2{\langle\textsl{g}^{2}GG\rangle}
\def\q2{\langle\bar{q}q\rangle}
\def\s2{\langle\bar{s}s\rangle}
\def\Gq{\langle\textsl{g}\bar{q}\sigma Gq\rangle}
\def\Gs{\langle\textsl{g}\bar{s}\sigma Gs\rangle}
\def\M{M_{B}}
\def\B{\mathcal{B}}

\begin{document}

\preprint{USTC-ICTS-09-01}

\title{~\\\vspace{2cm}Understanding light scalar meson by color-magnetic wavefunction in QCD sum rule

\vspace{1cm}}

\author{Yi Pang$^1$}\email{yipang@itp.ac.cn}
\author{Mu-Lin Yan$^2$}\email{mlyan@ustc.edu.cn}
\affiliation{$^1$Kavli Institute for Theoretical Physics China, Key
Laboratory of Frontiers in Theoretical Physics, Institute of
Theoretical Physics, Chinese Academy of Sciences, Beijing 100190,
P.R.China,\\$^2$Interdisciplinary Center of Theoretical Studies,
USTC, Hefei, Anhui 230026, P.R.China\vspace{2cm}
}%

\begin{abstract}
In this paper, we study the $0^{+}$ nonet mesons as tetraquark
 states
 with interpolating currents induced from the color-magnetic
 wavefunction. This wavefunction is the eigenfunction of effective color-magnetic Hamiltonian  with the
 lowest eigenvalue, meaning that the state depicted by this wavefunction is the most stable one and is most probable to be observed in
 experiments.
 Our approach can be recognized as
 determining interpolating currents dynamically.
 We perform an OPE calculation up to dimension eight condensates and find that the best QCD sum rule is achived when the current
 induced from the color-magnetic
 wavefunction is a proper mixture of the
 tensor and pseudoscalar diquark-antidiquark bound
 states. Compared with previous results, to sigma(600) and kappa(800), our results appear better, due to larger pole contribution.
  The direct instanton
 contribution are also considered,
  which yields a consistent result with
 previous OPE results. Finally, we also discuss the $\eta'$ problem as a possible six-quark state.
\end{abstract}

\pacs{12.38.Cy, 12.38.Lg, 12.39.Mk, 12.40.Yx}

\maketitle

\section{Introduction} In past decades, the question how to
validly interpret scalar mesons with their mass below 1 GeV
stimulated many discussions and controversies \cite{amsler}. In the
naive constituent quark model, they are expected to be $SU(3)_f$
nonet consisting of a quark and an antiquark, with one unit of
orbital excitation for positive parity. However, due to the fact
that the orbital excitation contributes energy about 0.5 GeV, it is
difficult to interpret their light mass as well as their mass
spectrum \cite{jaffe1}. Moreover, $a_0(980)$ and $f_0(980)$ couple
to $K\bar{K}$ channel strongly, which is in contradictory to the
prediction by naive $q\bar{q}$ mesons picture. This situation very
naturally leads to alternative interpretation about these mesons,
such as tetraquark states \cite{jaffe2,maiani,brito,wang,
zhu1,zhu2,zhu3,zhu4,lee1,lee2,lee3,lee4,Kojo,Matheus,Zhang,Latorre},
which were put forward many years ago in \cite{jaffe3,jaffe4}.
Recently, 't Hooft {\it et. al.} \cite{thooft} and authors of
\cite{schechter}
 found out new evidence from the instanton induced effective  Lagrangian, implying that the predominant
component of light scalar meson is tetraquark.

In 1977,  using the MIT bag model \cite{jaffe5}, Jaffe suggested the
existence of a light scalar nonet with masses below 1 GeV
\cite{jaffe3,jaffe4} . This nonet is composed by bound states of
diquark and antidiquark. The dominant interaction generating the
bound state is from one-gluon exchange which induces the following
effective Hamiltonian
\begin{equation}\label{H}
H_{eff}=-\widetilde{C}\sum_{i\neq j} (\lambda_i\cdot\lambda_j)
(\overrightarrow{\sigma}_i\cdot\overrightarrow{\sigma}_j),
\end{equation}
where $\widetilde{C}>0$ is the strength factor constant,
$\overrightarrow{\sigma}_i$ and $\lambda_i$ are $2\times 2$ Pauli
matrices and $3\times 3$ Gell-Mann color operators for the $i$th
quark. This is a simple generalization of the Breit spin-spin
interaction to include a similar color-color piece. It is also known
as ``color-magnetic" or ``color-spin" interaction of QCD, which was
first discussed in the pioneering work of De Rujula, Georgi, Glashow
\cite{DGG}. Hereafter, we will call the eigenstates of $H_{eff}$ as
color-magnetic eigenstates. The eigenfunctions and corresponding
eigenvalues of $H_{eff}$ for $q^2\bar{q}^2$ system (tetraquark) have
been presented in \cite{jaffe3,jaffe4}. In these work, the
eigenstate with the largest mass defect is
\begin{equation}\label{2}
|0^+, \underline{9}\rangle =0.972\; |0^+\underline{9}[1]\rangle
+0.233 \;|0^+\underline{9}[405]\rangle,
\end{equation}
with
\begin{equation}\label{5}
H_{eff}|0^+, \underline{9}\rangle =-43.36\widetilde{C} |0^+,
\underline{9}\rangle.
\end{equation}
 where $0^+$ stands for the $J^P$, $\underline{9}$ denotes flavor
$SU(3)_f$-nonet, and $\underline{9}[1]$ ($\underline{9}[405]$)
represents the nonet belonging to $[1]$-representation
($[405]$-representation) of color-spin $SU(6)_{CS}$. Explicitly,
they are
\begin{eqnarray}\label{3}
|0^+\underline{9}[1]\rangle &=& \sqrt{6\over 7}
|(6,3)\underline{\bar{3}};\; (\bar{6},\bar{3})\underline{3};\;
(1,1)\rangle + \sqrt{1\over 7}
|(\bar{3},1)\underline{\bar{3}};\; (3,1)\underline{3};\; (1,1)\rangle,\\
\label{4} |0^+\underline{9}[405]\rangle &=& \sqrt{1\over
7}|(6,3)\underline{\bar{3}};\; (\bar{6},\bar{3})\underline{3};\;
(1,1)\rangle- \sqrt{6\over 7} |(\bar{3},1)\underline{\bar{3}};\;
(3,1)\underline{3};\; (1,1)\rangle.
\end{eqnarray}
In the right hand side of above equations, there is state of
$|(6,3)\underline{\bar{3}};\; (\bar{6},\bar{3})\underline{3};\;
(1,1)\rangle $, where $(6,3)\underline{\bar{3}}$ indicates that the
diquark is in 6-dimension symmetric representation of color
$SU(3)_C$ with spin $S=1$ (so $2S+1=3$), and in 3-dimension
$\bar{3}$ representation of flavor $SU(3)_f$. While
$(\bar{6},\bar{3})\underline{3}$ means the antidiquark is in the
conjugate representation. And (1,1) means the bound state of diquark
and antidiquark is singlet both in color and spin. In the following,
without ambiguity, the diquark and antidiquark will be denoted
according to their $SU(3)_C$ representations. For example,
$\mathbf{6_c}$ diquark signifies the diquark's wavefunction is
$(6,3)\underline{\bar{3}}$. Similarly,
$|(\bar{3},1)\underline{\bar{3}};\; (3,1)\underline{3};\;
(1,1)\rangle$ is comprised of spin-0 $\mathbf{\bar{3}_c}$ diquark
and $\mathbf{3_c}$ antidiquark.

Basing on Eq. (\ref{5}), Jaffe claimed that the scalar tetraquarks
with masses below 1GeV exist and the color-spin part of their
wavefunctions can be described by $|0^+\underline{9}\rangle$.
Utilizing the latest data, Jaffe's statement could be roughly
checked for a visual comprehension. For instance, a data fit of
charmed baryons determines the constituent  quark masses
\cite{hogaasen1, hogaasen2, dy}:
\begin{equation}\label{6} m^{c}_u\approx m^{c}_d\approx 360{\rm
MeV}~~~m^{c}_s\approx 540{\rm MeV},
\end{equation}
where $c$ is the abbreviation of ``constituent''. The strength
factor constants related to the light quarks are
\begin{eqnarray}\label{7}
\widetilde{C}&\approx& \widetilde{C}_{qq}\approx 20{\rm
MeV},~~~~~{\rm with}\; \; q\in \{u,
d\},\\
\nonumber (\widetilde{C}_{qs} &=&15{\rm
MeV},~~~\widetilde{C}_{ss}=10{\rm MeV}).
\end{eqnarray}
Then, if we assume $\sigma(600)$ as one member of $0^+$-tetraquark
nonet, the mass of $\sigma(600)$ could be roughly estimated:
\begin{equation}\label{8}
m_{\sigma}\approx \langle \sum_i m_i^c-\widetilde{C}\sum_{i\; j}
(\lambda_i\cdot\lambda_j)
(\overrightarrow{\sigma}_i\cdot\overrightarrow{\sigma}_j)\rangle_\sigma\approx
4\times 360{\rm MeV} -43.36\times 20{\rm MeV}\approx 573{\rm MeV}.
\end{equation}
Obviously, Jaffe's  claim is reasonable, and  the underlying
dynamical consideration should be legitimate. Therefore, it is
interesting to study Jaffe's tetraquark in the framework of QCD sum
rule which relates the nonperturbative aspects of QCD to the
hadronic physics \cite{shifman, reinders}. In other words, we will
try to obtain a legitimate QCD sum rule for tetraquarks in terms of
their color-magnetic eigenfunctions.

QCD Sum Rule (SR) analysis for scalar nonet mesons as tetraquarks
has been widely discussed in the literature (e.g., see
\cite{brito,wang,
zhu1,zhu2,zhu3,zhu4,lee1,lee2,lee3,lee4,Kojo,Matheus,Zhang,Latorre}).
Since the correlator of tetraquark-type current operator for SR has
higher energy dimension than that of ordinary baryon-type one, the
operator product expansion (OPE) must be considered up to higher
dimensional operators (condensates) than ordinary baryons.
Technically, it has been widely accepted that the OPE contributions
from condensates of dimensions higher than eight are very small for
tetraquarks \cite{lee2}. To single scalar tetraquark current, it has
been shown in \cite{lee1} that the contributions from the dimension
eight condensates are unexpectedly large and become dominant in the
left hand sum rule. What is worse, their negative contributions
break down the physical meaning of the left hand sum rule. In order
to solve this problem, in \cite{lee2}, the authors demonstrated that
the current including equal weight of scalar and pseudoscalar
diquark-antidiquarks leads to a strong cancelation of the
contributions from dimension eight operators in the OPE, and then
gives a good sum rule. In \cite{zhu2},  by assuming mixing of single
tetraquark currents, the authors performed a SR analysis for
low-lying $0^+$-mesons as tetraquarks. However, by now, all work on
tetraquark SR has not considered a basic question that whether the
color-spin-flavor structures of the tetraquark-type currents in SR
are consistent with the color-magnetic hyperfine interaction
mechanism on tetraquarks. The aim of this paper is to pursue this
question.

The key point of this paper is that we think the interpolating
current used in SR should inherit a color-spin-flavor structure from
the color-magnetic wavefunction. This means that we treat a current
standing for linear combination of
$\mathbf{3_c}$-$\mathbf{\bar{3}_c}$ and
$\mathbf{6_c}$-$\mathbf{\bar{6}_c}$ tetraquarks as the SR
interpolating current. We emphasize that this combination or mixture
of $\mathbf{3_c}$-$\mathbf{\bar{3}_c}$ and
$\mathbf{6_c}$-$\mathbf{\bar{6}_c}$ tetraquarks is determined
dynamically by Eq. (\ref{5}) without any additional {\it ad hoc}
assumptions. Due to the non-relativistic nature of color-magnetic
interaction,
 it should be aware of that the induced mixture is specific to energy scale around 1GeV,
  which is mass scale of mesons we are interested in. In short, our method is based on the well established
concept that color-magnetic hyperfine interactions play a crucial
role in multiquark physics.

The strategy of the calculation is what follows. At the first step,
we will study the properties of the scalar tetraquark
$SU(3)_f$-nonet as color-magnetic eigenstate with the largest mass
defect in QCD sum rule by OPE expansion. With the method presented
in Section 2, we construct interpolating currents that can represent
the color-magnetic structure of tetraquark. Then utilizing these
currents, and following the standard procedure for tetraquark's OPE
calculations \cite{zhu1,zhu2,zhu3,zhu4,lee1,lee2,lee3,lee4}, we
obtain the contributions from the operators up to dimension eight.
Meanwhile, to achieve a reliable sum rule, we require that the pole
contributions should reach around $50\%$. Then we obtain $\sigma$
meson mass $(600\pm75)$MeV.

In addition, the instanton effects, in other words the topological
fluctuations of gluon fields, play an important role in the
structure of QCD vacuum \cite{schafer} and spectroscopy of
multiquark hadrons \cite{dorokhov, schafer1}. So they should be
taken into account in the SR calculations. Combining the
contribution from OPE and instanton, we obtain $\sigma$ mass about
720 MeV close to previous OPE results. At this stage, a complete sum
rule description of $0^{+}$ nonet meson has been obtained by us,
including both the OPE and instanton effects.

The paper is organized as follows. In Section II, we will deduce the
interpolating currents for $0^{+}$ tetraquarks from their
color-magnetic wavefunctions. In Section III, the analytic results
of OPE calculation based on previous currents will be presented,
followed by the numerical results. In Section IV, the single direct
instanton contribution will be considered.  In Section V, we
summarize the results briefly and make a speculation on the
extension of our method to study mesons with 6 quarks (Fermi-Yang
meson). In appendix, we will list some necessary formulas of
spectral functions and correlators.

 \section{Interpolating current for Jaffe tetraquark }
\label{S:Introduction} { Substituting Eqs. (\ref{3}) and (\ref{4})
into (\ref{2}), we obtain the expression of the color-magnetic
wavefunction for Jaffe's $0^{+}$ tetraquark nonet meson as follows
\begin{equation}\label{wavefunction}
    |0^{+},9\rangle =
    0.988|(6,3)\underline{\overline{3}};(\overline{6},\overline{3})\underline{3};(1,1)\rangle
    +
    0.157|(\overline{3},1)\underline{\overline{3}};(3,1)\underline{3};(1,1)\rangle.
\end{equation}
The elements for $|0^{+},9\rangle$ are $\mathbf{6_c}$,
$\mathbf{\bar{3}_c}$  diquarks and $\mathbf{{\bar{6}}_c}$,
${\mathbf{3}_c}$ anti-diquarks. Generally, the composite operator
for a diquark with certain structure of color, flavor and spin is
\begin{equation}\label{10}
    \sum_{\{a\leftrightarrow b\},\{i\leftrightarrow j\}} (-1)^{P_c}(-1)^{P_f}
    q^{(i)\;T}_a C
    \Gamma q^{(j)}_b,
\end{equation}
where $\{a,b\}$ and $\{i,j\}$ are color and flavor indices of quarks
respectively. Specifically, $q^{(1)}_a=u_a,\; q^{(2)}_a=d_a,
\;q^{(3)}_a=s_a$. $C$ is the charge conjugation operator, and
$\Gamma$ is Dirac matrix determined by the spin of the system.
$(-1)^{P_c}$ and $(-1)^{P_f}$ reflect the parities of the diquark's
color and flavor wavefunctions respectively. As for wavefunctions
being symmetric in color or flavor, $P_{c}=0$, or $P_f=0$, and for
anti-symmetric ones, $P_c=1$ or $P_f=1$. Notation
$\{a\leftrightarrow b\},\{i\leftrightarrow j\}$ represent the color
and flavor permutations respectively. Since
$|(6,3)\underline{\overline{3}};(\overline{6},\overline{3})\underline{3};(1,1)\rangle$
signifies that the diquark and anti-diquark are symmetric in color
and anti-symmetric in flavor, the composite operator of
$\mathbf{6_c}$ diquark  can be written as
\begin{equation}\label{11}
 q^{(i)\;T}_a C \Gamma q^{(j)}_b -q^{(j)\;T}_a C \Gamma q^{(i)}_b +q^{(i)\;T}_b C \Gamma
 q^{(j)}_a -q^{(j)\;T}_b C \Gamma q^{(i)}_a.
\end{equation}
 In the
non-relativistic limit of diquark bispinor $q^TC\Gamma q$, spin-1
requires that
\begin{equation}\label{12}
\Gamma=\{ \sigma^{\mu\nu},~ \gamma^\mu,~\gamma^\mu\gamma^5\}~~~~{\rm
with}\;\;\sigma^{\mu\nu}={i\over 2} (\gamma^\mu\gamma^\nu-\gamma^\nu
\gamma^\mu).
\end{equation}
Then, inserting (\ref{12}) into (\ref{11}), we obtain all possible
composite operators for $\mathbf{6_c}$ spin-1 diquark expressed as
below,
\begin{eqnarray}
\nonumber Q_T^{(ij)}(6)&\equiv&{1\over 2\sqrt{2}}(q^{(i)\;T}_a C
\sigma^{\mu\nu} q^{(j)}_b -q^{(j)\;T}_a C \sigma^{\mu\nu} q^{(i)}_b
+q^{(i)\;T}_b C \sigma^{\mu\nu}
 q^{(j)}_a -q^{(j)\;T}_b C \sigma^{\mu\nu} q^{(i)}_a)\\
\label{13}&=&{1\over \sqrt{2}}(q^{(i)\;T}_a C \sigma^{\mu\nu}
q^{(j)}_b -q^{(j)\;T}_a
C \sigma^{\mu\nu} q^{(i)}_b) ,\\
\nonumber Q_A^{(ij)}(6)&\equiv&{1\over 2\sqrt{2}}(q^{(i)\;T}_a C
\gamma^\mu q^{(j)}_b -q^{(j)\;T}_a C \gamma^\mu q^{(i)}_b
+q^{(i)\;T}_b C \gamma^\mu
 q^{(j)}_a -q^{(j)\;T}_b C \gamma^\mu q^{(i)}_a)\\
\label{14}&=&{1\over \sqrt{2}}(q^{(i)\;T}_a C \gamma^\mu q^{(j)}_b
-q^{(j)\;T}_a
C \gamma^\mu q^{(i)}_b) ,\\
\nonumber Q_B^{(ij)}(6)&\equiv&{1\over 2\sqrt{2}}(q^{(i)\;T}_a C
\gamma^\mu\gamma^5 q^{(j)}_b -q^{(j)\;T}_a C \gamma^\mu\gamma^5
q^{(i)}_b +q^{(i)\;T}_b C \gamma^\mu\gamma^5
 q^{(j)}_a -q^{(j)\;T}_b C \gamma^\mu\gamma^5 q^{(i)}_a)\\
 \label{15}&=&0.
\end{eqnarray}
where   $1/(2\sqrt{2})$ is a widely adopted normalization. Likewise,
the composite operators of $\mathbf{\bar{6}_c}$ spin-1 antidiquark
are
\begin{eqnarray}
\label{16} \overline{Q}_T^{(ij)}(6)&=& {1\over
\sqrt{2}}(\bar{q}^{(i)}_a \sigma_{\mu\nu} C\bar{q}^{(j)\;T}_b
-\bar{q}^{(j)}_a
 \sigma_{\mu\nu} C\bar{q}^{(i)\;T}_b) ,\\
\label{17} \overline{Q}_A^{(ij)}(6)&=&{1\over
\sqrt{2}}(\bar{q}^{(i)}_a \gamma_\mu C \bar{q}^{(j)\;T}_b
-\bar{q}^{(j)}_a
 \gamma_\mu C\bar{q}^{(i)\;T}_b ),\\
\label{18} \overline{Q}_B^{(ij)}(6)&=&0.
\end{eqnarray}
Because
$|(\overline{3},1)\underline{\overline{3}};(3,1)\underline{3};(1,1)\rangle$
means that the diquark and antidiquark are anti-symmetric in color,
spin and flavor. The composite operators for $\mathbf{\bar{3}_c}$
spin-0 diquarks belonging to representation
$(\bar{3},1)\underline{\overline{3}}$ of $SU(6)_{cs}\times SU(3)_f$
 are the following ones,
\begin{eqnarray}
\label{19} {Q}_S^{(ij)}(3)&=& {1\over \sqrt{2}}({q}^{(i)\;T}_a C
\gamma^5{q}^{(j)}_b -{q}^{(j)\;T}_a
 C \gamma^5{q}^{(i)}_b) ,\\
\label{20} {Q}_P^{(ij)}(3)&=& {1\over \sqrt{2}}({q}^{(i)\;T}_a C
{q}^{(j)}_b -{q}^{(j)\;T}_a
 C {q}^{(i)}_b).
\end{eqnarray}
On the other hand, the composite operators of $\mathbf{{3}_c}$
spin-0 antidiquarks belonging to the conjugate representation
 are
\begin{eqnarray}
\label{21} \overline{Q}_S^{(ij)}(3)&=& {1\over
\sqrt{2}}(\bar{q}^{(i)}_a \gamma^5 C\bar{q}^{(j)\;T}_b
-\bar{q}^{(j)}_a
 \gamma^5 C\bar{q}^{(i)\;T}_b), \\
\label{22} \overline{Q}_P^{(ij)}(3)&=&{1\over
\sqrt{2}}(\bar{q}^{(i)}_a
 C \bar{q}^{(j)\;T}_b -\bar{q}^{(j)}_a
  C\bar{q}^{(i)\;T}_b ).
\end{eqnarray} }

For the time being, we can express the composite operators related
to
$|(6,3)\underline{\overline{3}};(\overline{6},\overline{3})\underline{3};(1,1)\rangle$
as
\begin{eqnarray}
\nonumber T_6^{\{ij\}\{lm\}}&\equiv&Q_T^{(ij)}(6)
\overline{Q}_T^{(lm)}(6)\\
\label{23} &=& q^{(i)\;T}_a C \sigma^{\mu\nu} q^{(j)}_b
 \bar{q}^{(m)}_a
 \sigma_{\mu\nu} C\bar{q}^{(l)\;T}_b + q^{(i)\;T}_b C \sigma^{\mu\nu}
 q^{(j)}_a
 \bar{q}^{(l)}_a
 \sigma_{\mu\nu} C\bar{q}^{(m)\;T}_b ,\\
 \nonumber A_6^{\{ij\}\{lm\}}&\equiv&Q_A^{(ij)}(6)
\overline{Q}_A^{(lm)}(6)\\
\label{24} &=& q^{(i)\;T}_a C \gamma^\mu q^{(j)}_b
 \bar{q}^{(m)}_a
 \gamma_\mu C\bar{q}^{(l)\;T}_b + q^{(i)\;T}_b C \gamma^\mu
 q^{(j)}_a
 \bar{q}^{(l)}_a
 \gamma_\mu C\bar{q}^{(m)\;T}_b ,
\end{eqnarray}
where $T$, $A$ represent ``tensor'' and ``axial vector''
respectively. These notations lie with how the diquark and
anti-diquark operators vary under Lorentz transformation. In terms
of Eqs. (\ref{19})-(\ref{22}), the composite operators corresponding
to $|(\bar{3},1)\underline{\overline{3}};(3,
1)\underline{3};(1,1)\rangle$ are
\begin{eqnarray}
\nonumber S_3^{\{ij\}\{lm\}}&\equiv&Q_S^{(ij)}(3)
\overline{Q}_S^{(lm)}(3)\\
\label{25} &=& \epsilon_{abc}\epsilon_{ab'c'}q^{(i)\;T}_b C \gamma^5
q^{(j)}_c
 \bar{q}^{(m)}_{b'}
 \gamma^5 C\bar{q}^{(l)\;T}_{c'}, \\
\nonumber P_3^{\{ij\}\{lm\}}&\equiv&Q_P^{(ij)}(3)
\overline{Q}_P^{(lm)}(3)\\
\label{26} &=& \epsilon_{abc}\epsilon_{ab'c'}q^{(i)\;T}_b C
q^{(j)}_c
 \bar{q}^{(m)}_{b'}
  C\bar{q}^{(l)\;T}_{c'},
\end{eqnarray}
where $S$, $P$ stand for ``scalar'' and ``pseudoscalar''
respectively. Following Jaffe, $\{\sigma,\;f_0,\;a_+,\;\kappa \}$
are assumed as $ 0^+$-$SU(3)_f$ nonet tetraquarks. For $\sigma$,
since its flavor content is $\{ud\}\{\bar{u}\bar{d}\}$, by Eqs.
(\ref{23}) and (\ref{24}), the operators corresponding to
$|(6,3)\underline{\overline{3}};(\overline{6},\overline{3})\underline{3};(1,1)\rangle
$ of $\sigma$ are
\begin{eqnarray}
 \label{27}  T_{6}^{\sigma}&\equiv&T_6^{\{ud\}\{\bar{u}\bar{d}\}}=u^{T}_{a}C\sigma^{\mu\nu}d_{b}\bar{d}_{a}\sigma_{\mu\nu}C\bar{u}^{T}_{b}+u^{T}_{b}C\sigma^{\mu\nu}d_{a}\bar{u}_{a}\sigma_{\mu\nu}C\bar{d}^{T}_{b},\\
 \label{28}  A_{6}^{\sigma}&\equiv&A_6^{\{ud\}\{\bar{u}\bar{d}\}}=u^{T}_{a}C\gamma^{\mu}d_{b}\bar{d}_{a}\gamma_{\mu}C\bar{u}^{T}_{b}+u^{T}_{b}C\gamma^{\mu}d_{a}\bar{u}_{a}\gamma_{\mu}C\bar{d}^{T}_{b}.
\end{eqnarray}
By Eqs. (\ref{25}) and (\ref{26}), the operators corresponding to
$|(\overline{3},1)\underline{\overline{3}};(3,1)\underline{3};(1,1)\rangle
$ of $\sigma$ are
\begin{eqnarray}
\label{29}  S_{3}^{\sigma} &\equiv&S_3^{\{ud\}\{\bar{u}\bar{d}\}}= \epsilon_{abc}\epsilon_{ab^{\pri}c^{\pri}}u^{T}_{b}C\gamma^{5}d_{c}\bar{u}_{b^{\pri}}\gamma^{5}C\bar{d}^{T}_{c^{\pri}},\\
\label{30}
P_{3}^{\sigma}&\equiv&P_3^{\{ud\}\{\bar{u}\bar{d}\}}=\epsilon_{abc}\epsilon_{ab^{\pri}c^{\pri}}u^{T}_{b}Cd_{c}\bar{u}_{b^{\pri}}C\bar{d}^{T}_{c^{\pri}}.
\end{eqnarray}
Similarly, for $f_{0}$, because of its flavor content
$\frac{1}{\sqrt{2}}(\{us\}\{\bar{u}\bar{s}\}+\{ds\}\{\bar{d}\bar{s}\})$,
the results are
\begin{eqnarray}
\nonumber    T_{6}^{f_{0}}\equiv T_6^{\frac{1}{\sqrt{2}}(\{us\}\{\bar{u}\bar{s}\}+\{ds\}\{\bar{d}\bar{s}\})}&=&\frac{1}{\sqrt{2}}(u^{T}_{a}C\sigma^{\mu\nu}s_{b}\bar{s}_{a}\sigma_{\mu\nu}C\bar{u}^{T}_{b}+u^{T}_{b}C\sigma^{\mu\nu}s_{a} \bar{u}_{a}\sigma_{\mu\nu}C\bar{s}^{T}_{b})  \\
\label{31}                && + \frac{1}{\sqrt{2}}(d^{T}_{a}C\sigma^{\mu\nu}s_{b}\bar{s}_{a}\sigma_{\mu\nu}C\bar{d}^{T}_{b}+d^{T}_{b}C\sigma^{\mu\nu}s_{a}\bar{d}_{a}\sigma_{\mu\nu}C\bar{s}^{T}_{b}),\\
\nonumber  A_{6}^{f_{0}}\equiv A_6^{\frac{1}{\sqrt{2}}(\{us\}\{\bar{u}\bar{s}\}+\{ds\}\{\bar{d}\bar{s}\})}&=&\frac{1}{\sqrt{2}}(u^{T}_{a}C\gamma^{\mu}s_{b}\bar{s}_{a}\gamma_{\mu}C\bar{u}^{T}_{b}+u^{T}_{b}C\gamma^{\mu}s_{a}\bar{u}_{a}\gamma_{\mu}C\bar{s}^{T}_{b}) \\
 \label{32}
 &&\frac{1}{\sqrt{2}}(d^{T}_{a}C\gamma^{\mu}s_{b}\bar{s}_{a}\gamma_{\mu}C\bar{d}^{T}_{b}+d^{T}_{b}C\gamma^{\mu}s_{a}\bar{d}_{a}\gamma_{\mu}C\bar{s}^{T}_{b}),
\end{eqnarray}
\begin{eqnarray}
\label{33} S_{3}^{f_{0}} \equiv
S_3^{\frac{1}{\sqrt{2}}(\{us\}\{\bar{u}\bar{s}\}+\{ds\}\{\bar{d}\bar{s}\})}&=&\hskip-0.1in
\frac{1}{\sqrt{2}}\epsilon_{abc}(\epsilon_{ab^{\pri}c^{\pri}}u^{T}_{b}C\gamma^{5}s_{c}\bar{u}_{b^{\pri}}\gamma^{5}C\bar{s}^{T}_{c^{\pri}}
+\epsilon_{ab^{\pri}c^{\pri}}d^{T}_{b}C\gamma^{5}s_{c}\bar{d}_{b^{\pri}}\gamma^{5}C\bar{s}^{T}_{c^{\pri}}), \\
\label{34}  P_{3}^{f_{0}}\equiv
S_3^{\frac{1}{\sqrt{2}}(\{us\}\{\bar{u}\bar{s}\}+\{ds\}\{\bar{d}\bar{s}\})}&=&\frac{1}{\sqrt{2}}\epsilon_{abc}(\epsilon_{ab^{\pri}c^{\pri}}u^{T}_{b}Cs_{c}\bar{u}_{b^{\pri}}C\bar{s}^{T}_{c^{\pri}}
+\epsilon_{ab^{\pri}c^{\pri}}d^{T}_{b}Cs_{c}\bar{d}_{b^{\pri}}C\bar{s}^{T}_{c^{\pri}}).
\end{eqnarray}

 The results for $a_{+}~(\{us\}\{\bar{d}\bar{s}\})$, $\kappa~
 (\{ud\}\{\bar{d}\bar{s}\})$ are the following ones,
\begin{eqnarray}
  \label{35}  T_{6}^{a_{+}}&\equiv&T_6^{(\{us\}\{\bar{d}\bar{s}\})}=u^{T}_{a}C\sigma^{\mu\nu}s_{b}\bar{d}_{a}\sigma_{\mu\nu}C\bar{s}^{T}_{b}+u^{T}_{b}C\sigma^{\mu\nu}s_{a}\bar{d}_{a}\sigma_{\mu\nu}C\bar{s}^{T}_{b},  \\
 \label{36}  A_{6}^{a_{+}}&\equiv&A_6^{(\{us\}\{\bar{d}\bar{s}\})}=u^{T}_{a}C\gamma^{\mu}s_{b}\bar{d}_{a}\gamma_{\mu}C\bar{s}^{T}_{b}+u^{T}_{b}C\gamma^{\mu}s_{a}\bar{d}_{a}\gamma_{\mu}C\bar{s}^{T}_{b}, \\
\label{37}   S_{3}^{a_{+}} &\equiv&S_3^{(\{us\}\{\bar{d}\bar{s}\})}= \epsilon_{abc}\epsilon_{ab^{\pri}c^{\pri}}u^{T}_{b}C\gamma^{5}s_{c}\bar{d}_{b^{\pri}}\gamma^{5}C\bar{s}^{T}_{c^{\pri}}, \\
 \label{38}
 P_{3}^{a_{+}}&\equiv&P_3^{(\{us\}\{\bar{d}\bar{s}\})}=\epsilon_{abc}\epsilon_{ab^{\pri}c^{\pri}}u^{T}_{b}Cs_{c}\bar{d}_{b^{\pri}}C\bar{s}^{T}_{c^{\pri}}.
\end{eqnarray}
\begin{eqnarray}{}
 \label{39}   T_{6}^{\kappa}&\equiv&T_6^{(\{ud\}\{\bar{d}\bar{s}\})}=u^{T}_{a}C\sigma^{\mu\nu}d_{b}\bar{s}_{a}\sigma_{\mu\nu}C\bar{d}^{T}_{b}+u^{T}_{b}C\sigma^{\mu\nu}d_{a}\bar{s}_{a}\sigma_{\mu\nu}C\bar{d}^{T}_{b},  \\
 \label{40}  A_{6}^{\kappa}&\equiv&A_6^{(\{ud\}\{\bar{d}\bar{s}\})}=u^{T}_{a}C\gamma^{\mu}d_{b}\bar{s}_{a}\gamma_{\mu}C\bar{d}^{T}_{b}+u^{T}_{b}C\gamma^{\mu}d_{a}\bar{s}_{a}\gamma_{\mu}C\bar{d}^{T}_{b}, \\
 \label{41}  S_{3}^{\kappa} &\equiv&S_3^{(\{ud\}\{\bar{d}\bar{s}\})}= \epsilon_{abc}\epsilon_{ab^{\pri}c^{\pri}}u^{T}_{b}C\gamma^{5}d_{c}\bar{s}_{b^{\pri}}\gamma^{5}C\bar{d}^{T}_{c^{\pri}}, \\
 \label{42}
 P_{3}^{\kappa}&\equiv&P_3^{(\{ud\}\{\bar{d}\bar{s}\})}=\epsilon_{abc}\epsilon_{ab^{\pri}c^{\pri}}u^{T}_{b}Cd_{c}\bar{s}_{b^{\pri}}C\bar{d}^{T}_{c^{\pri}}.
\end{eqnarray}

Subsequently, from above results and basing on Eq.
(\ref{wavefunction}), we get the desired all possible simplest
interpolating currents for tetraquark $|0^+,9\rangle$ as follows
\begin{equation}\label{current1}
 \nonumber   J^{X}_{1}= \alpha T_{6}^{X}+\beta S_{3}^{X},
\end{equation}
\begin{equation}\label{current2}
  \nonumber   J^{X}_{2}= \alpha T_{6}^{X}+\beta P_{3}^{X},
\end{equation}
\begin{equation}\label{current3}
  \nonumber   J^{X}_{3}= \alpha A_{6}^{X}+\beta S_{3}^{X},
\end{equation}
\begin{equation}\label{current4}
    J^{X}_{2}= \alpha A_{6}^{X}+\beta P_{3}^{X},
\end{equation}
where $X$ can signifies $\sigma, \kappa, a_{+}$ and $f_{0}$, with
$\alpha=0.988$ and $\beta=0.157$. We notice that some indispensable
contents of the best mixed current in \cite{zhu2} disappear here.
The reason is that they are forbidden by requiring the wavefunction
of diquark to be anti-symmetrized \cite{jaffe3,jaffe4}.

\section{QCD sum rule analysis without instanton contribution}
\subsection{General formulas for QCD sum rule}
In sum rule analysis, we usually consider two-point correlation
functions:
\begin{equation}\label{corre1}
    \Pi(q^{2})\equiv i\int d^{4}x e^{iqx}\langle0|{\rm T}
    J(x)J^{\dagger}(0)|0\rangle,
\end{equation}
where $J$ is an interpolating current for the tetraquark. We compute
$\Pi(q^{2})$ up to certain order in the expansion, which is matched
with a hadronic parametrization to extract information of hadron
properties. At hadron level, we express the correlation function in
the form of dispersion relation with a spectral function:
\begin{equation}\label{corre2}
  \Pi(q^{2})= \int^{\infty}_{0} \frac{\rho(s)}{s-q^{2}-i\epsilon}ds,
\end{equation}
where
\begin{eqnarray}\label{spectral}
 \nonumber  \rho(s) &=&\pi\sum_{n}\delta(s-M_{n}^{2})\langle0|
    J(x)|n\rangle \langle n|J^{\dagger}(0)|0\rangle, \\
   &=&2\pi f_{X}^{2}m_{X}^{8}\delta(s-M_{X}^{2})+\rm higher~states,
\end{eqnarray}
with the convention
\begin{equation}\label{}
\langle0|
    J(x)|S_{i}\rangle =\sqrt{2}f_{i}m_{i}^{4}.
\end{equation}
The sum rule analysis is then performed after Borel transforming
both sides of Eqs. (\ref{corre1}) and (\ref{corre2}),
\begin{equation}\label{lhs1}
    \Pi^{(\rm all)}(M_{B}^{2})=\B_{M_{B}^{2}}\Pi(q^{2})=\int^{\infty}_{0}
    e^{-s/M_{B}^{2}}\rho(s)ds.
\end{equation}
Usually, evaluating $\rho(s)$ by OPE or some other methods, then
from Eq. (\ref{lhs1}), one obtains the left hand sum rule (LHS). On
the other hand, inserting Eq. (\ref{spectral}) into Eq.
(\ref{lhs1}), one derives the right hand sum rule (RHS). By
definition,
\begin{equation}\label{rhs}
\Pi_{\rm RHS}(M_{B}^{2})=2\pi
f_{X}^{2}m_{X}^{8}e^{-m_{X}^{2}/M_{B}^{2}}.
\end{equation}
The LHS and RHS are supposed to be equal, so we obtain
\begin{equation}\label{LandR}
    \int^{S_{0}}_{0}
    e^{-s/M_{B}^{2}}\rho(s)ds=2\pi f_{X}^{2}m_{X}^{8}e^{-m_{X}^{2}/M_{B}^{2}}.
\end{equation}
In above expressions, we have chosen a finite threshold $S_{0}$ to
exclude the contribution from the continuum. Differentiating Eq.
(\ref{LandR}) with respect to $\frac{1}{M_{B}^{2}}$, and dividing it
by Eq. (\ref{LandR}), finally we obtain the physical mass
\begin{equation}\label{lhs2}
    M_{X}^{2}=\frac{\int^{S_{0}}_{0}
    e^{-s/M_{B}^{2}}s\rho(s)ds}{\int^{S_{0}}_{0}
    e^{-s/M_{B}^{2}}\rho(s)ds}.
\end{equation}
 In the following, we
study both Eqs. (\ref{lhs1}) and (\ref{lhs2}) as functions of Borel
mass $M_{B}$ and threshold $S_{0}$.
\subsection{OPE calculation for $0^{+}$ nonet as Jaffe tetraquark}
The $\sigma$-correlator can be expressed as follows:
\begin{eqnarray}
\nonumber    \Pi^{\sigma}(q^{2}) &=&i\int d^{4}xe^{iq\cdot x}\langle0|TJ^{\sigma}(x)J^{\sigma\dag}(0)|0\rangle\\
   &=& \alpha^{2}\Pi^{\sigma {\rm OPE}}_{A,A}+\beta^{2}\Pi^{\sigma {\rm OPE}}_{B,B}+\alpha\beta(\Pi^{\sigma {\rm OPE}}_{A,B}+\Pi^{\sigma
{\rm OPE}}_{B,A}).
\end{eqnarray}
where $J^{\sigma}=\alpha A+\beta B$ represents any one of the four
possible currents in Eq. (\ref{current4}), $A$ represents the
composite operator related to $\mathbf{6_c}$-$\mathbf{\bar{6}_c}$,
and $B$ is that associated with $\mathbf{3_c}$-$\mathbf{\bar{3}_c}$.
$\Pi_{A,B}$ is the correlator between $A$-type content and $B$-type
content. In this section, we will first compute the spectral
functions for the correlators through OPE expansion, then insert
these results into the Eq. (\ref{lhs1}) to obtain the Borel
transformed correlators. In the process of calculating OPE, we use
the following propagators for quarks \cite{zhu1}, which contain all
the necessary terms for computing tetraquark spectral functions.
\begin{eqnarray}\label{propagator1}
 \nonumber iS_{q}^{ab}(x)&\equiv&\langle0|T[q_{a}(x)\bar{q}_{b}(0)]|0\rangle\\
\nonumber
 &=&\frac{i\delta^{ab}}{2\pi^{2}x^{4}}\hat{x}+\frac{i}{32\pi^{2}}\frac{\lambda^{n}_{ab}}{2}\textsl{g}_{c}G^{n}_{\mu\nu}\frac{1}{x^{2}}(\sigma^{\mu\nu}\hat{x}+\hat{x}\sigma^{\mu\nu})-\frac{\delta^{ab}}{12}\langle\bar{q}q\rangle+\frac{\delta^{ab}x^{2}}{192}\langle\textsl{g}_{c}\bar{q}\sigma
               Gq\rangle-\frac{\delta^{ab}m_{q}}{4\pi^{2}x^{2}}\\
 &&+\frac{i\delta^{ab}m_{q}}{48}\langle\bar{q}q\rangle\hat{x}+\frac{i\delta^{ab}m_{q}^{2}}{8\pi^{2}x^{2}}\hat{x}~~~~~~~{\rm with}~ q\in \{u, d\}.
\end{eqnarray}
\begin{eqnarray}\label{propagator2}
 \nonumber iS_{s}^{ab}(x)&\equiv&\langle0|T[s_{a}(x)\bar{s}_{b}(0)]|0\rangle\\
\nonumber
 &=&\frac{i\delta^{ab}}{2\pi^{2}x^{4}}\hat{x}+\frac{i}{32\pi^{2}}\frac{\lambda^{n}_{ab}}{2}\textsl{g}_{c}G^{n}_{\mu\nu}\frac{1}{x^{2}}(\sigma^{\mu\nu}\hat{x}+\hat{x}\sigma^{\mu\nu})-\frac{\delta^{ab}}{12}\langle\bar{s}s\rangle+\frac{\delta^{ab}x^{2}}{192}\langle\textsl{g}_{c}\bar{s}\sigma
               Gs\rangle-\frac{\delta^{ab}m_{s}}{4\pi^{2}x^{2}}\\
 &&+\frac{i\delta^{ab}m_{s}}{48}\langle\bar{s}s\rangle\hat{x}+\frac{i\delta^{ab}m_{s}^{2}}{8\pi^{2}x^{2}}\hat{x}.
\end{eqnarray}
 Actually, OPE computation for tetraquarks is rather long,
but it can be performed analytically. A convenient formulation  for
performing this calculation has been presented in \cite{zhu1,zhu2}.
The MATHMATICA with FEYNCALC \cite{feynman} may be helpful for
computation. In the following, we use the notations and formulations
in \cite{zhu1,zhu2}. We have performed the OPE calculation for
spectral functions up to dimension eight, which is up to the
constant ($s^0$) term of $\rho(s)$. During the calculations, we have
assumed the vacuum is saturated for higher dimension operators, such
as
$\langle0|\bar{q}q\bar{q}q|0\rangle\sim\langle0|\bar{q}q|0\rangle^2$.
After finishing the OPE calculation, we obtain the following results
for $\sigma$ meson,
\begin{eqnarray}
\nonumber \rho^{\sigma \rm OPE}_{T,T}
&=&\frac{s^{4}}{1280}-\frac{m_{q}^{2}}{16\pi^{6}}s^{3}+(\frac{21m_{q}^{4}}{16\pi^{6}}+\frac{\langle\bar{q}q\rangle
 m_{q}}{2\pi^{4}}+\frac{11\G2}{768})s^{2}-(\frac{9m_{q}^{6}}{2\pi^{6}}+\frac{15\q2m_{q}^{3}}{\pi^{4}}+\frac{11\G2m_{q}^{2}}{64\pi^{6}})s\\
 &&+(\frac{9m_{q}^{8}}{4\pi^{6}}+\frac{18\q2m_{q}^{5}}{\pi^{4}}+\frac{11\G2m_{q}^{4}}{64\pi^{6}}-\frac{3\Gq
  m_{q}^{3}}{\pi^{4}}+\frac{30\q2^{2}m_{q}^{2}}{\pi^{2}}+\frac{11\G2\q2m_{q}}{48\pi^{4}}),\\
 \nonumber&&\\
\nonumber \rho^{\sigma \rm OPE}_{S,S}&=& \frac{s^{4}}{61440\pi^{6}}-\frac{m_{q}^{2}s^{3}}{1536\pi^{6}}+(\frac{3m_{q}^{4}}{256\pi^{6}}-\frac{m_{q}\q2}{96\pi^{4}}+\frac{\G2}{6144\pi^{6}})s^{2}-(\frac{3m_{q}^{6}}{64\pi^{6}}+\frac{\G2m_{q}^{2}}{1024\pi^{6}}+\frac{\Gq m_{q}}{32\pi^{4}}\\
  && -\frac{\q2^{2}}{12\pi^{2}})s+(\frac{3m_{q}^{8}}{64\pi^{6}}+\frac{\G2m_{q}^{4}}{512\pi^{6}}+\frac{\Gq m_{q}^{3}}{16\pi^{4}}-\frac{m_{q}^{2}\q2^{2}}{24\pi^{2}}-\frac{\G2\q2m_{q}}{384\pi^{4}}+\frac{\Gq\q2}{12\pi^{2}}),\\
 \nonumber   &&\\
 \rho^{\sigma \rm OPE}_{T,S}&=&\rho^{\sigma
 \rm OPE}_{S,T}=-\frac{\G2}{1024\pi^{6}}s^{2}+\frac{3\G2m_{q}^{2}}{256\pi^{6}}s-(\frac{3\G2m_{q}^{4}}{256\pi^{6}}+\frac{\G2\q2m_{q}}{64\pi^{4}}),\\
\nonumber&&\\
\nonumber  \rho^{\sigma \rm OPE}_{P,P}&=&
\frac{s^{4}}{61440\pi^{6}}-\frac{m_{q}^{2}s^{3}}{512\pi^{6}}+(\frac{11m_{q}^{4}}{256\pi^{6}}+\frac{m_{q}\q2}{32\pi^{4}}+\frac{\G2}{6144\pi^{6}})s^{2}-(\frac{9m_{q}^{6}}{64\pi^{6}}+\frac{5\q2m_{q}^{3}}{8\pi^{4}}+\frac{3\G2m_{q}^{2}}{1024\pi^{6}}\\~~~~~~
\nonumber&&-\frac{\Gq
m_{q}}{32\pi^{4}}+\frac{\q2^{2}}{12\pi^{2}})s+(\frac{3m_{q}^{8}}{64\pi^{6}}+\frac{3\q2m_{q}^{5}}{4\pi^{4}}+\frac{\G2m_{q}^{4}}{512\pi^{6}}-\frac{3\Gq
   m_{q}^{3}}{16\pi^{4}}+\frac{31m_{q}^{2}\q2^{2}}{24\pi^{2}}\\
   &&+\frac{\G2\q2m_{q}}{128\pi^{4}}-\frac{\Gq\q2}{12\pi^{2}}),\\
   \nonumber&&\\
 \rho^{\sigma \rm OPE}_{T,P}&=&\rho^{\sigma
 \rm OPE}_{P,T}=-\frac{\G2}{512\pi^{6}}s^{2}+\frac{3\G2m_{q}^{2}}{128\pi^{6}}s-(\frac{3\G2m_{q}^{4}}{128\pi^{6}}+\frac{\G2\q2m_{q}}{32\pi^{4}}),\\
\nonumber&& \end{eqnarray}
\begin{eqnarray}
  \nonumber \rho^{\sigma \rm OPE}_{A,A}&=&\frac{s^{4}}{7680\pi^{6}}-\frac{m_{q}^{2}s^{3}}{128\pi^{6}}+(\frac{5m_{q}^{4}}{32\pi^{6}}+\frac{\G2}{3072\pi^{6}})s^{2}-(\frac{9m_{q}^{6}}{16\pi^{6}}+\frac{5\q2m_{q}^{3}}{4\pi^{4}}+\frac{9\G2m_{q}^{2}}{512\pi^{6}}+\frac{\Gq m_{q}}{8\pi^{4}}\\
   && +\frac{\q2^{2}}{3\pi^{2}})s+(\frac{3m_{q}^{8}}{8\pi^{6}}+\frac{3\q2m_{q}^{5}}{2\pi^{4}}+\frac{5\G2m_{q}^{4}}{256\pi^{6}}+\frac{7m_{q}^{2}\q2^{2}}{3\pi^{2}}+\frac{\G2\q2m_{q}}{64\pi^{4}}+\frac{\Gq\q2}{3\pi^{2}}),\\
\nonumber   &&\\
 \rho^{\sigma \rm OPE}_{A,S}&=&\rho^{\sigma
 \rm OPE}_{S,A}=-\frac{3\G2m_{q}^{2}}{1024\pi^{6}}s+\frac{\G2\q2m_{q}}{64\pi^{4}},\\
\rho^{\sigma \rm OPE}_{A,P}&=&\rho^{\sigma \rm OPE}_{P,A}=0.
\end{eqnarray}
In above equations, $\q2$ is a dimension $d=3$ quark condensate;
$\G2$ is a dimension $d=4$ gluon condensate; $\Gq$ is a dimension
$d=5$ mixed condensate; the strong coupling constant takes its value
at energy scale about 1 GeV, that is the energy scale we are
interested in. Long distance bulk properties of physical vacuum are
effectively parameterized in these vacuum expectation values. At
present, according to Eq. (43), we can make use of above spectral
functions to generate correlator of each kind interpolating current
belonging to $\sigma$. These correlators will be the starting point
of numerical calculation in the next section.

In order to prevent the long listing of formulas for spectral
functions from obscuring the conceptual content, we will put the
necessary spectral functions of $\kappa$, $a_+$ and $f_0$ into the
appendix.

\subsection{Numerical analysis of QCD sum rule for OPE contribution}
For numerical calculations, we use the following values of
condensates~\cite{Yang:1993bp,Narison:2002pw,Gimenez:2005nt,Jamin:2002ev,Ioffe:2002be,Ovchinnikov:1988gk,Yao:2006px}:
%
\begin{eqnarray}\label{condensates}
\nonumber &&\langle\bar qq \rangle=-(0.240 \mbox{ GeV})^3\, ,
\\
\nonumber &&\langle\bar ss\rangle=-(0.8\pm 0.1)\times(0.240 \mbox{
GeV})^3\, ,
\\
\nonumber &&\langle g_s^2GG\rangle =(0.48\pm 0.14) \mbox{ GeV}^4\, ,
\\ \nonumber && m_u =m_d = m_q=0.1\times2.4^{-3} \mbox{ GeV}\, ,
\\
 &&m_s(1\mbox{ GeV})=125 \pm 20 \mbox{ MeV}\, ,
\\
\nonumber && \langle g_s\bar q\sigma G
q\rangle=-M_0^2\times\langle\bar qq\rangle\, ,
\\
\nonumber &&M_0^2=(0.8\pm0.2)\mbox{ GeV}^2\, .
\end{eqnarray}

\begin{figure}[ht]
\begin{center}
\vspace{0ex}
\includegraphics[width=.42\textwidth]{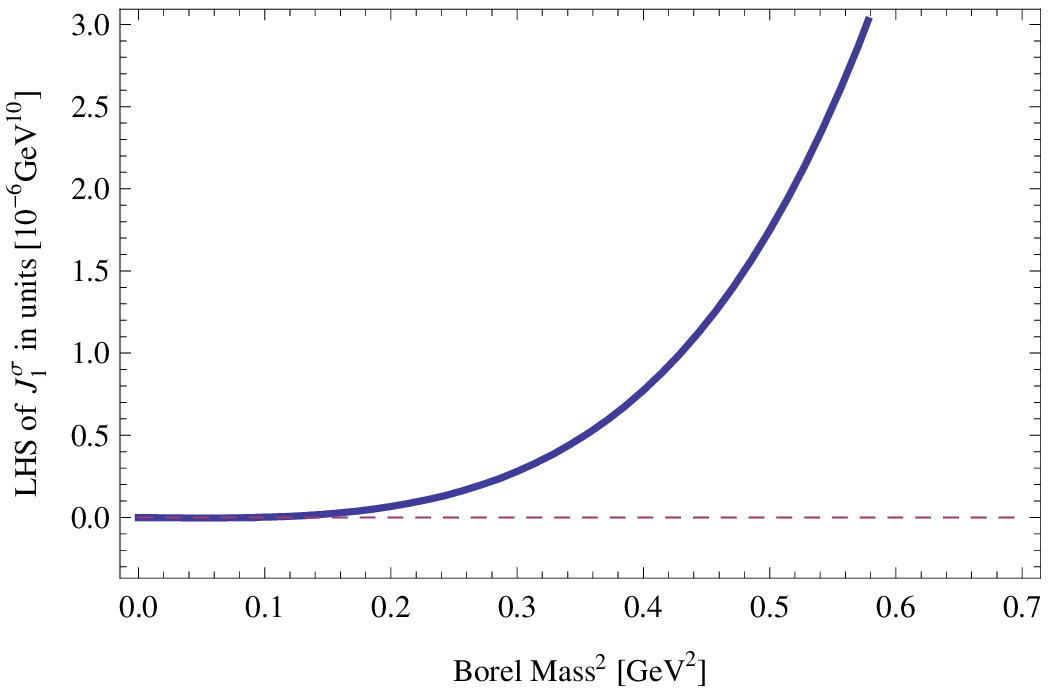}
\includegraphics[width=.42\textwidth]{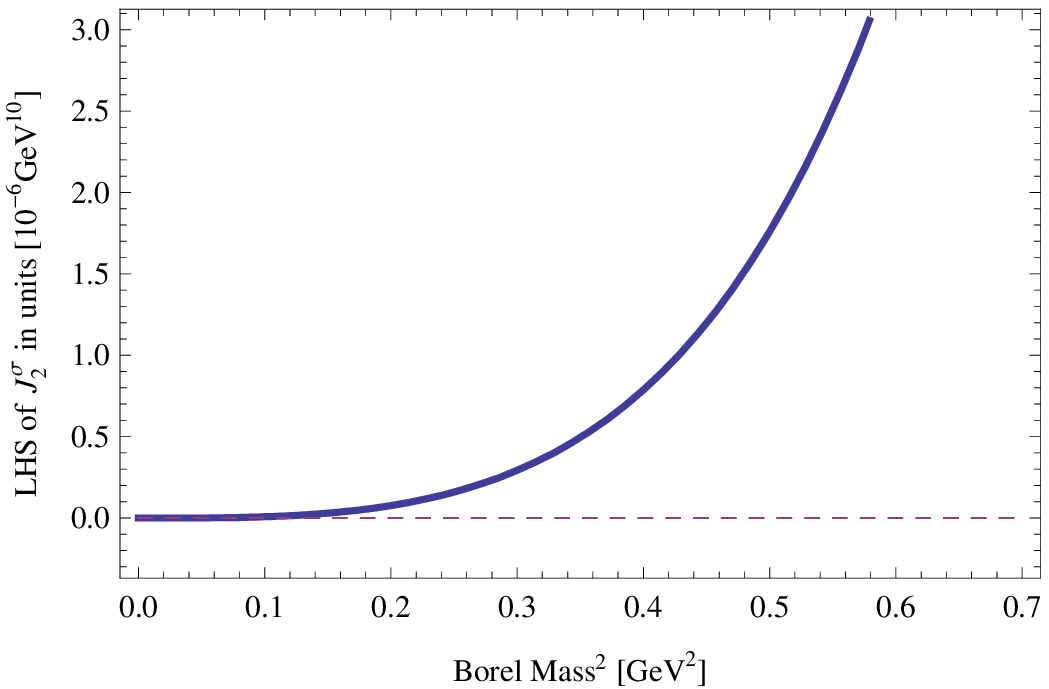}
\includegraphics[width=.42\textwidth]{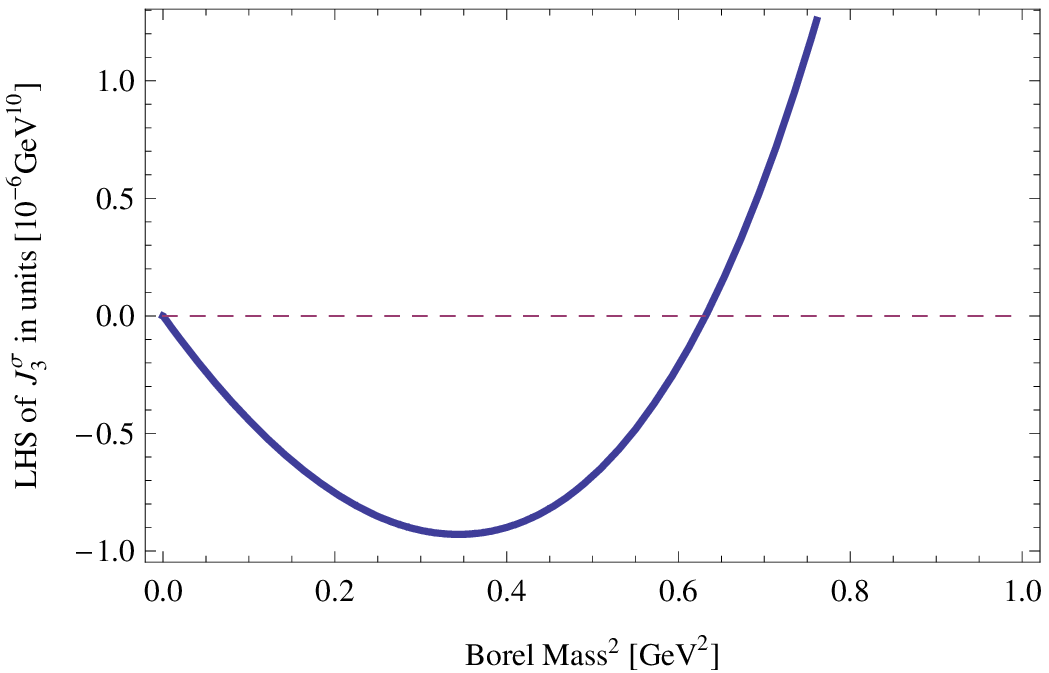}
\includegraphics[width=.42\textwidth]{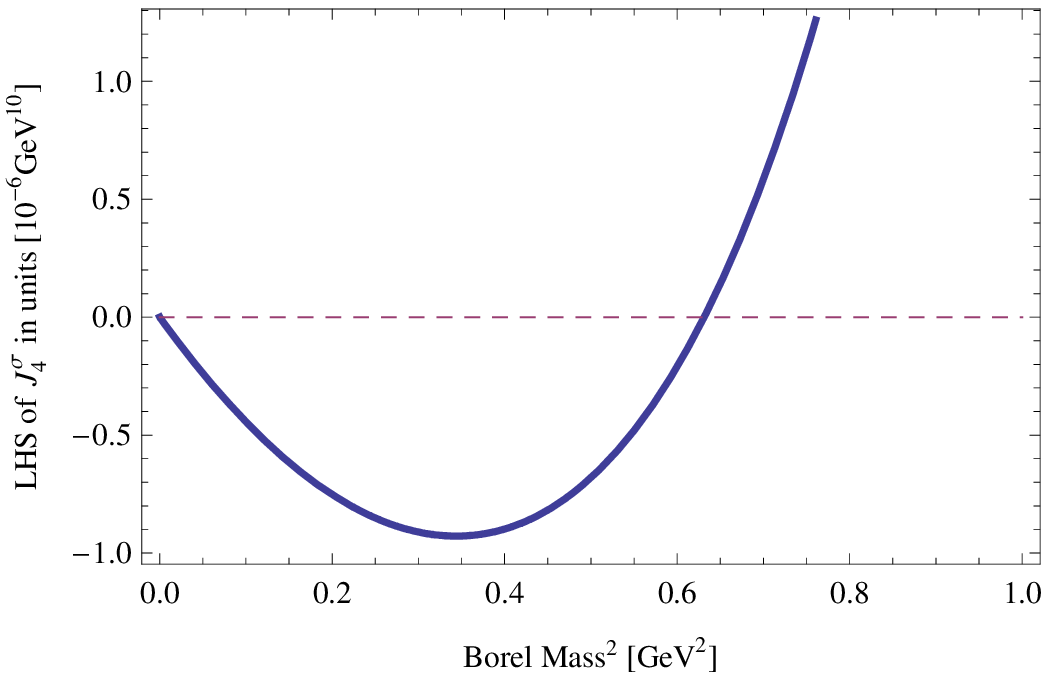}
\end{center}
\caption{\label{Ls}LHS of four interpolating currents of $\sigma$
meson, as a functions of Borel mass squared, with $s_{0}$=infinity,
in units of $\rm GeV^{10}$.}
\end{figure}

Figure 1 shows the LHS of four possible interpolating currents of
the $\sigma$ meson, as a function of Borel mass squared, in the case
of infinite threshold. From the definition of Eq. (\ref{lhs1}), the
LHS should be positive quantities. However, in practical
calculations, the positivity may not be necessarily realized due to
the insufficient convergence of OPE calculations. In our case, from
Figure. \ref{Ls}, we see that current $J_1^{\sigma}$ and current
$J_2^{\sigma}$ show better convergence than current $J_3^{\sigma}$
and current $J_4^{\sigma}$.

\begin{figure}[hbt]
\begin{center}
\vspace{0ex}
\includegraphics[width=.45\textwidth]{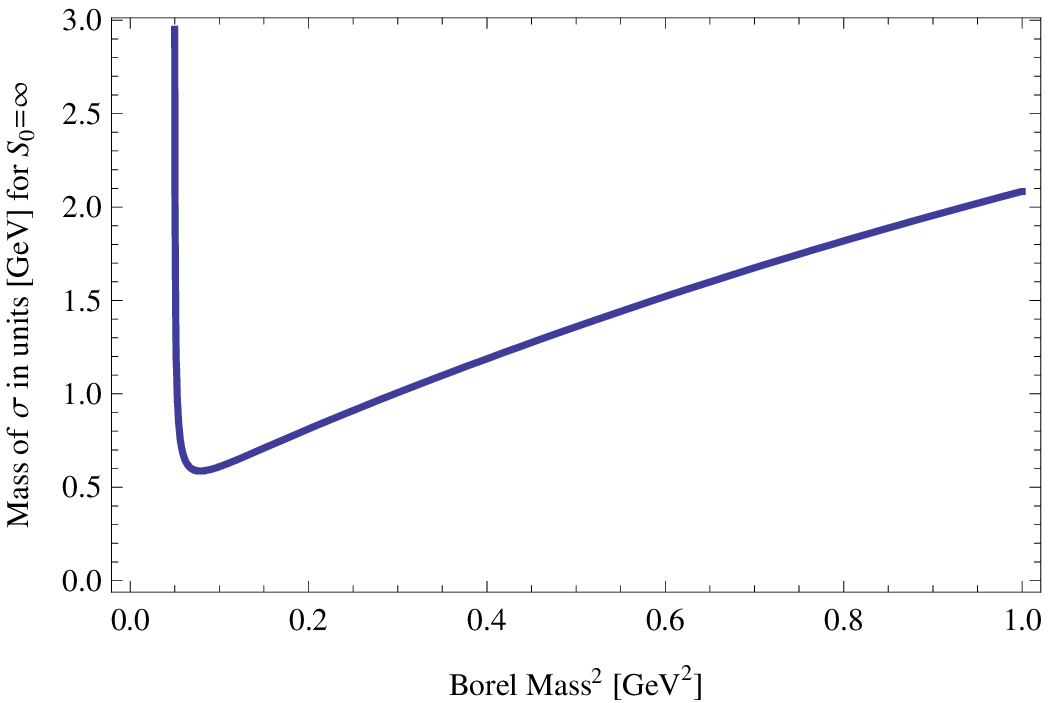}
\includegraphics[width=.45\textwidth]{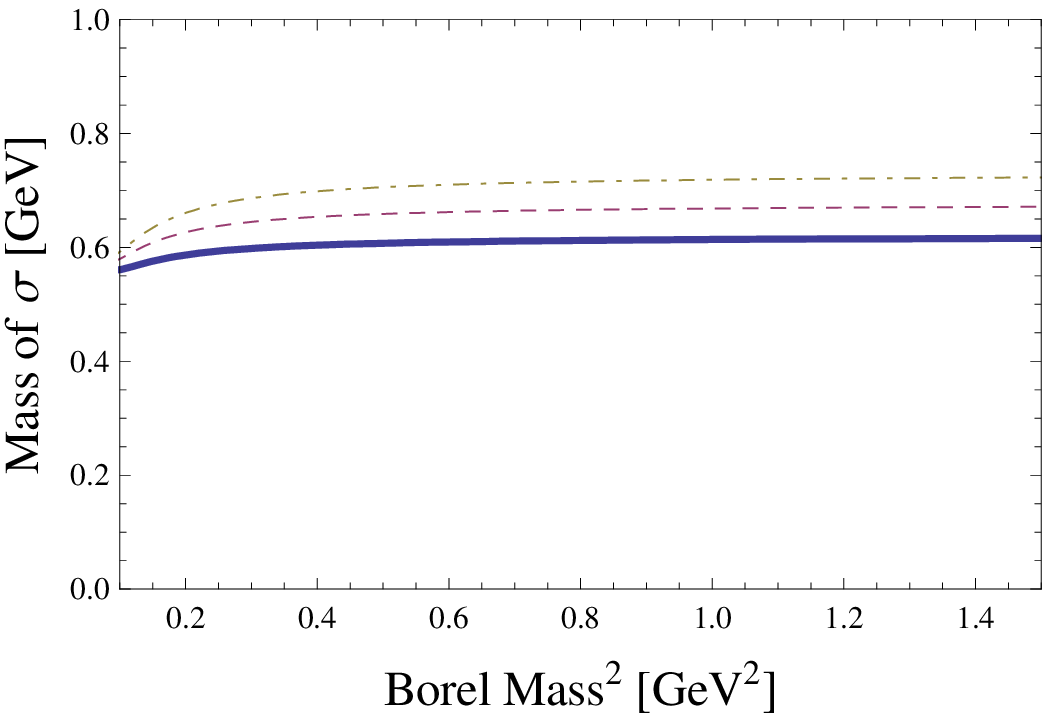}
\end{center}
\caption{\label{Ms} Mass of $\sigma$ is illustrated as function of
Borel Mass squared. The left figure is in the case of infinite
threshold, while the right one is in cases of finite thresholds. The
results corresponding to $s_0$ =0.5, 0.6, 0.7 $\rm GeV^2$ are
represented by a solid line, a dashed line and a dot-dashed line
respectively.}
\end{figure}

To find the current with the best convergence, we have to refer to
their Borel transformed correlators in numerical expressions, which
are:
\begin{eqnarray}\label{spectrum-sigma}
  \nonumber  \Pi^{\sigma(\rm all)}_{1}&=& 1.9\times10^{-5}\M^{10}-1.9\times10^{-8}\M^{8}+9.5\times10^{-6}\M^{6}+3.7\times10^{-8}\M^{4}-8.5\times10^{-8}\M^{2}, \\
  \nonumber \Pi^{\sigma(\rm all)}_{2}&=&1.9\times10^{-5}\M^{10}-2.0\times10^{-8}\M^{8}+9.5\times10^{-6}\M^{6}-4.2\times10^{-8}\M^{4}-2.1\times10^{-8}\M^{2},  \\
 \nonumber  \Pi^{\sigma(\rm all)}_{3}&=& 3.2\times10^{-6}\M^{10}-2.5\times10^{-9}\M^{8}+1.6\times10^{-6}\M^{6}-6.2\times10^{-6}\M^{4}-5.1\times10^{-6}\M^{2} ,\\
 \Pi^{\sigma(\rm all)}_{4}&=&3.2\times10^{-6}\M^{10}-2.5\times10^{-9}\M^{8}+1.6\times10^{-6}\M^{6}+6.2\times10^{-6}\M^{4}-5.1\times10^{-6}\M^{2}.
\end{eqnarray}
From these expressions,  it is obvious that current $J_2^{\sigma}$
shows the best convergence behavior, so we will utilize current
$J_2^{\sigma}$ to compute the physical mass of $\sigma$. We first
choose an infinite threshold to estimate the mass as the traditional
sum rule has done \cite{reinders}. In Figure \ref{Ms}, we exhibit
the behavior of the mass of $\sigma$ meson as the function of
$M_{B}$ for infinite and finite $s_{0}$. In  traditional sum rule,
if the mass as a function of $M_{B}$, has a wide minimum, then the
minimum value of mass function can be perceived as the real mass of
the state. From Figure \ref{Ms}, we observed that $M_{\sigma}$ as a
function of $M_B^2$ indeed has a minimum with $M_{\sigma(\rm
min)}=0.59 {\rm~GeV}$ at $M_{B}^{2}=0.079 {\rm~GeV}^{2}$. At this
value of Borel mass, the correlation function $\Pi^{\sigma(\rm
all)}_{2}=3\times10^{-9}~{\rm GeV}^{10}$, so the positivity of LHS
is kept. Although $M_{\sigma(\rm min)}$ is very close to the
experimental center value $\langle M_{\sigma}\rangle\sim0.6 \rm
~GeV$, the minimum is not wide enough as required. Therefore, to
obtain an acceptable result, we have to adopt finite thresholds
scheme \cite{zhu1,zhu2,zhu3,zhu4} to repeat the process of computing
mass. The results for some values of threshold are presented in the
right part of Figure 2. We notice that when the mass becomes weakly
dependent on $M_{B}$, the value of mass is around 0.6 GeV. But we
also find that as the threshold increases, the mass will increase
too. This may be due to the fact that $\sigma$ is a broad resonance
state. So there must be some criteria to help us dictate which value
of mass is the most believable one. Combining the points of view
adopted by\cite{zhu2,Kojo,Matheus} on judging when an acceptable sum
rule is arrived, we postulate the following criteria.

1. The Borel transformed correlation function $\Pi(M_{B}^{2})$
should show a good positivity for almost all values of Borel mass.
This is usually related the convergence of LHS.

2. The physical mass should depend weakly on the value of Borel mass
in a wide region. In other words, there should be a Borel window.

\begin{figure}[ht]
\begin{center}
\vspace{0ex}
\includegraphics[width=.45\textwidth]{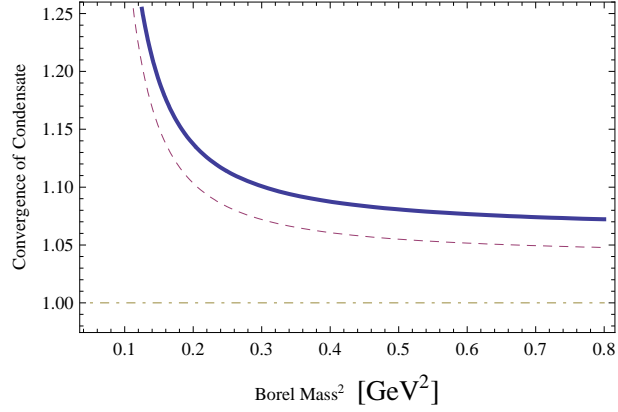}
\end{center}
\caption{\label{sc}convergence of OPE series of spectral function
related to current$J^{\sigma}_2$ for $s_0=0.6 \rm {GeV}^2$.}
\end{figure}

3. OPE convergence. This is a strong constraint to the lower bound
of the $M_{B}^2$ region. OPE series converge better for higher
values of $M_{B}^2$, so that requiring a good convergence sets a
lower limit to $M_{B}^2$. To current $J^{\sigma}_2$, we find such a
lower limit of $M_{B}^2$ in the following. We first rewrite the
spectral function corresponding to $J^{\sigma}_2$ as,
\begin{equation}\label{OPEseria}
    \rho_{\sigma}^{(\rm {OPE})}=\Sigma_{n=0}^{4}c^{(8-2n)}s^n=\Sigma_{n=0}^{4}\rho^n,
\end{equation}
where $c^{(8-2n)}$ denotes the operators of mass dimension $(8-2n)$,
$\rho^n\equiv c^{(8-2n)}s^n$. From Eqs. (55)-(62), we learn that
terms $\rho^{(3,4)}$ are perturbative contributions denoted as
$\rho^{(pert)}$, in other words, they do not contain condensate.
Remaining terms represent contributions from operators of dimension
4, 6 and 8. These terms are dominated by condensates including the
non-perturbative effect, denoted by $\rho^{(2)}$, $\rho^{(1)}$,
$\rho^{(0)}$ respectively. In Fig. \ref{sc}, we present the relative
contribution of $\rho^{(2)}$, $\rho^{(1)}$, $\rho^{(0)}$ to the
total spectral function $\rho^{({\rm OPE})}_{\sigma}$.
The thick line denotes [$\int^{0.6}_0
(\rho^{(pert)}+\rho^{(2)})e^{-s/M_B^2}ds/\int^{0.6}_0 \rho^{(\rm
OPE)}e^{-s/M_B^2}ds$], the dashed line signifies [$\int^{0.6}_0
(\rho^{(pert)}+\rho^{(2)}+\rho^{(1)})e^{-s/M_B^2}ds/\int^{0.6}_0
\rho^{(\rm OPE)}e^{-s/M_B^2}ds$], the dashed doted line represents
[$\int^{0.6}_0
(\rho^{(pert)}+\rho^{(2)}+\rho^{(1)}+\rho^{(0)})e^{-s/M_B^2}ds/\int^{0.6}_0
\rho^{(\rm OPE)}e^{-s/M_B^2}ds$=1]. We see that, for $M_B^2>0.2\rm
{GeV}^2$, the addition of a subsequent term in expansion
(\ref{OPEseria}), brings the curve closer to an asymptotic value
(which is normalized to 1). Furthermore, the changes in this curve
become smaller with increasing dimension. Thus, for
$s_0=0.6\rm{GeV}^2$, the convergence is satisfied by $M_B^2>0.2\rm
{GeV}^2$. For $s_0=0.5,~0.7, ~0.8\rm{GeV}^2$, convergence limits
$M_B^2>0.2,~0.3,~0.4\rm {GeV}^2$, respectively.

 4. For a given threshold, the pole contribution should be sufficient large.
 By choosing suitable Borel mass, this can be satisfied. Since the Borel
transformation suppresses the contributions from $s_0>M_B^2$, small
value of $M_B^2$ are preferred to suppress the continuum
contributions. But $M_B^2$ cannot be arbitrarily small, or it will
spoil previous three requirements. To $\sigma$, we have found such
optimal values of $M_B^2$ for different thresholds. We list the
corresponding pole contributions in Table I. The pole contribution
is defined as
\begin{equation}\label{eq_pole}
\mbox{Pole contribution} \equiv \frac{ \int^{s_0}_0 e^{-s/M_B^2}
\rho(s)ds }{\int^\infty_0 e^{-s/M_B^2} \rho(s)ds}\, .
\end{equation}
\begin{table}[hbt]
\caption{Pole contributions of various threshold.}
\begin{center}
\begin{tabular}{c|c|c|c|c}
 \hline  $s_{0}~({\rm GeV}^{2})$ & 0.5 & 0.6 & 0.7 & 0.8\\
 \hline   $M_B^{2}~({\rm GeV}^{2})$ & 0.2 & 0.2  & 0.3 & 0.4\\
 \hline   Pole (\%) & 40 & 52 & 35 & 25\\
 \hline    $M_{\sigma}$ (GeV) & 0.6 & 0.6 & 0.7  & 0.75\\
  \hline
\end{tabular}
\end{center}
\end{table}

 From this table, we can extract following information that when threshold
 changes from $0.5~{\rm GeV}^{2}$ to $0.8~{\rm
 GeV}^{2}$, the pole contribution will vary from 40\% to 25\% correspondingly, but reaches its maximum 52\% at $M_B^2$=0.2 ${\rm GeV}^2$, when $s_0=0.6{\rm GeV}^2$. That the pole contribution reaches 52\% implies that a good sum rule
 has
 been obtained. We get
 \begin{equation}\label{sigmaM}
    m_{\sigma}=(600\pm75)\rm
    {MeV},~~~~~~~\rm{with~Pole~contribution}(52\%),
 \end{equation}
where $(\pm 75)$ MeV originates from the error of condensates (see
Eq. \ref{condensates}). It is remarkable that the Pole contribution
is larger than that given in \cite{zhu2}, where the Pole
contribution is below 30\%.

\begin{figure}[ht]
\begin{center}
\vspace{0ex}
\includegraphics[width=.42\textwidth]{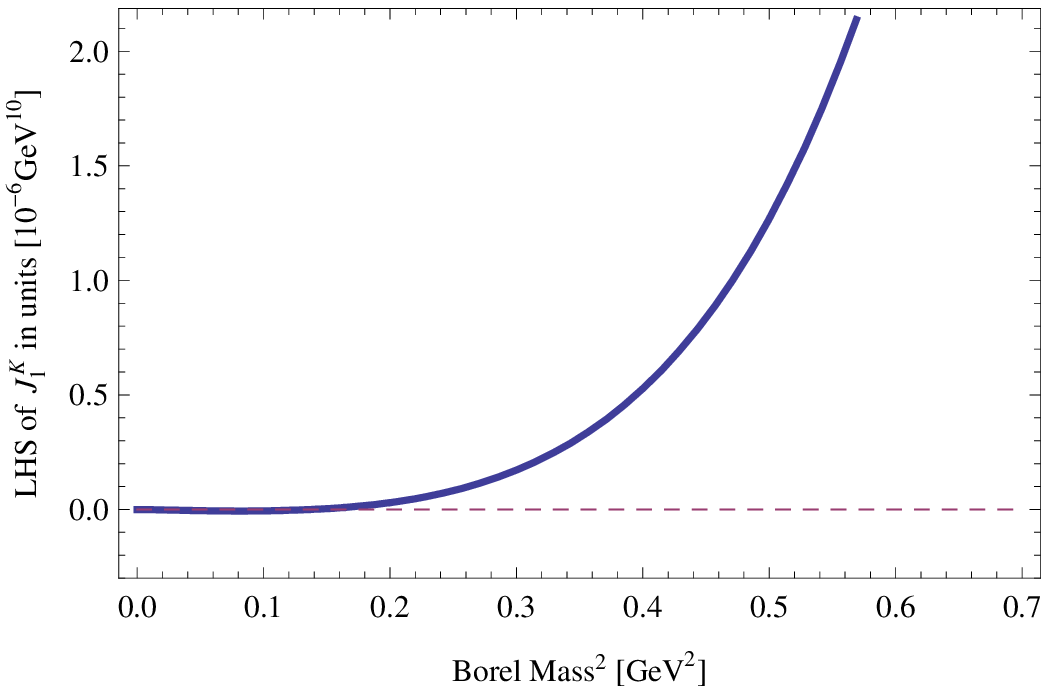}
\includegraphics[width=.42\textwidth]{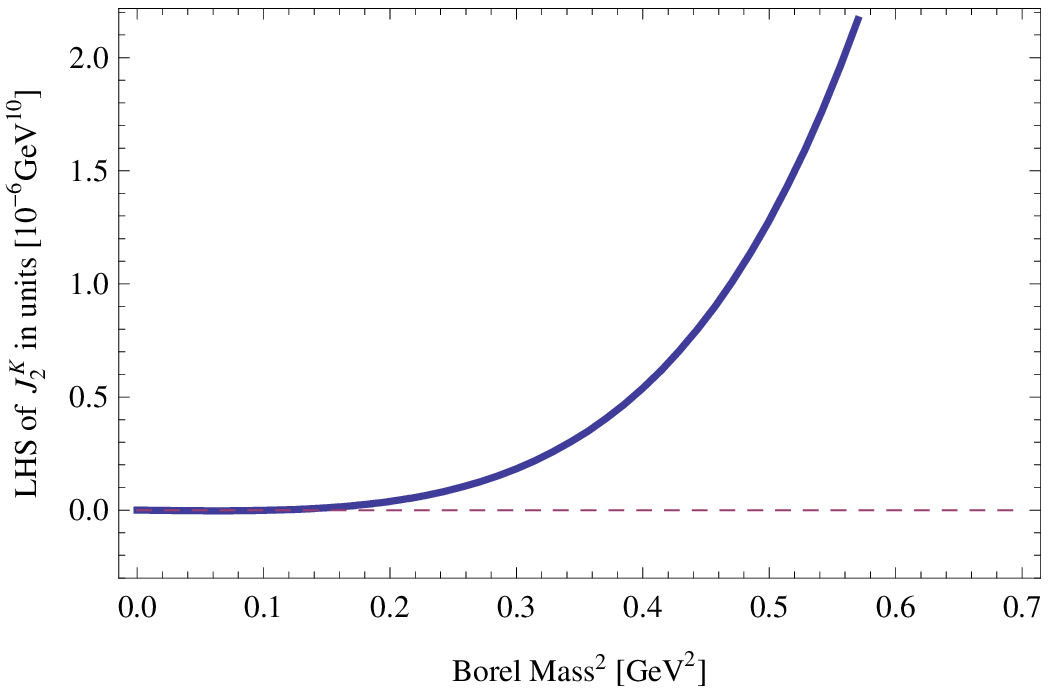}
\includegraphics[width=.42\textwidth]{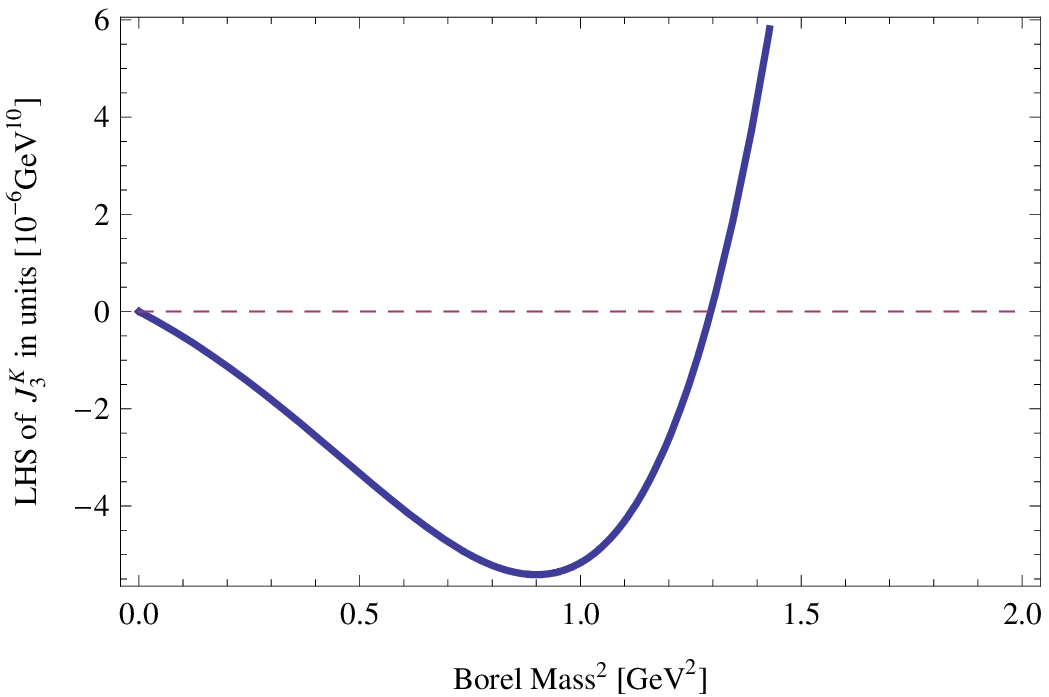}
\includegraphics[width=.42\textwidth]{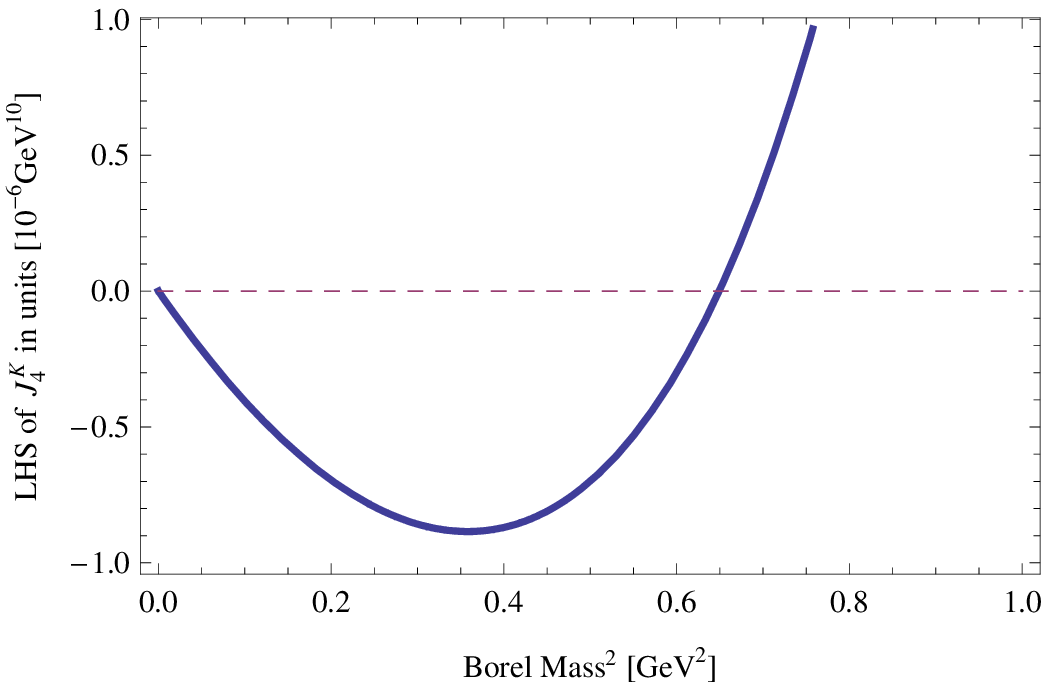}
\end{center}
\caption{\label{kappa1} LHS of $\kappa$ meson as functions of Borel
mass squared with $s_{0}$=infinity in units of $\rm GeV^{10}$. }
\end{figure}

\begin{figure}[ht]
\begin{center}
\vspace{0ex}
\includegraphics[width=.45\textwidth]{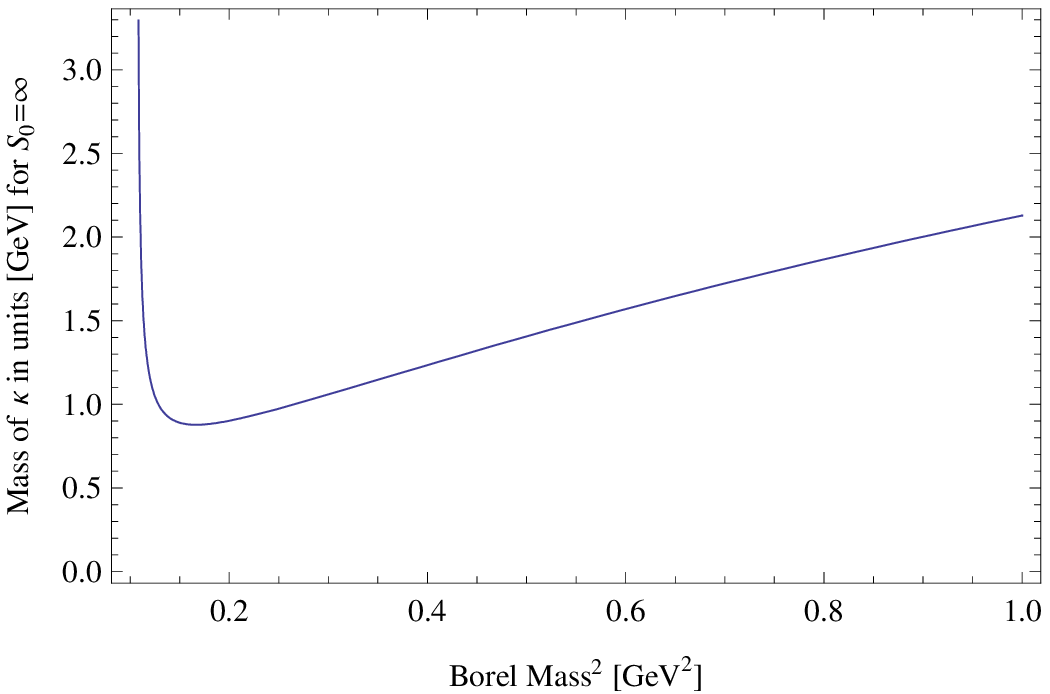}
\includegraphics[width=.45\textwidth]{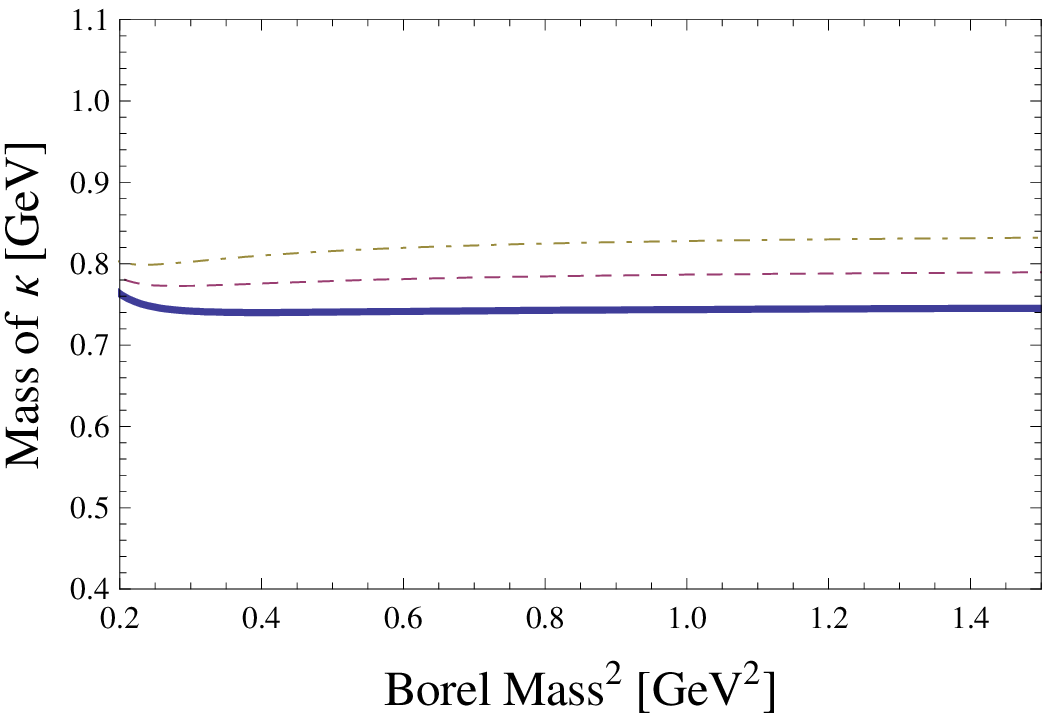}
\end{center}
\caption{\label{Mk}Mass of $\kappa$ is illustrated as function of
Borel Mass squared. The left figure is in the case of infinite
threshold, while the right one is in cases of finite thresholds. The
results corresponding to $s_0$ =0.7, 0.8, 0.9 $\rm GeV^2$ are
represented by a solid line, a dashed line and a dot-dashed line
respectively. }
\end{figure}

\begin{figure}[ht]
\begin{center}
\vspace{0ex}
\includegraphics[width=.45\textwidth]{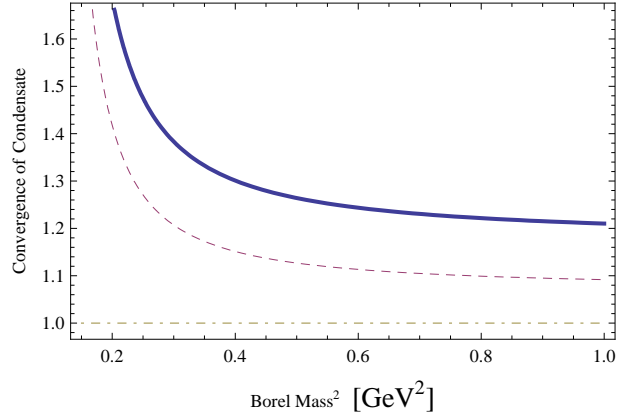}
\end{center}
\caption{\label{kc}convergence of OPE series of spectral function
related to current$J^{\kappa}_2$ for $s_0=0.9 \rm {GeV}^2$.}
\end{figure}

 Applying the same analysis to meson $\kappa$, the LHS of four possible
interpolating currents of $\kappa$ can be found in Figure
\ref{kappa1}, with threshold value $s_{0}$ being infinity.
 The corresponding numerical expressions are listed below:
\begin{eqnarray}\label{spectrum-kappa}
  \nonumber  \Pi^{\kappa(\rm all)}_{1}&=& 1.9\times10^{-5}\M^{10}-1.2\times10^{-6}\M^{8}+6.7\times10^{-6}\M^{6}-1.3\times10^{-7}\M^{4}-1.2\times10^{-7}\M^{2}, \\
  \nonumber \Pi^{\kappa(\rm all)}_{2}&=&1.9\times10^{-5}\M^{10}-1.2\times10^{-6}\M^{8}+6.7\times10^{-6}\M^{6}-1.9\times10^{-7}\M^{4}-5.7\times10^{-8}\M^{2} , \\
 \nonumber  \Pi^{\kappa(\rm all)}_{3}&=& 3.2\times10^{-6}\M^{10}-1.9\times10^{-7}\M^{8}+1.7\times10^{-6}\M^{6}-5.2\times10^{-6}\M^{4}-4.6\times10^{-6}\M^{2}, \\
  \Pi^{\kappa(\rm all)}_{4}&=&3.2\times10^{-6}\M^{10}-1.9\times10^{-7}\M^{8}+1.7\times10^{-6}\M^{6}+5.2\times10^{-6}\M^{4}-4.6\times10^{-6}\M^{2}.
\end{eqnarray}
From Figure \ref{kappa1} and above expressions, we notice that
current $J_2^{\kappa}$, which is a proper mixture between tensor and
pseudoscalar contents, is the best interpolating current. By setting
the threshold to be infinity, we obtain an estimation for the mass
of $\kappa$. As shown in Figure \ref{Mk}, $M_{\kappa}$ as a function
of $M_B$ has a minimum with $M_{\kappa(\rm min)}=0.90 {\rm~GeV}$ at
$M_{B}^{2}=0.2 {\rm~GeV}^{2}$. At this value of Borel mass, the
correlation function $\Pi^{\kappa(\rm
all)}_{2}=1.6\times10^{-7}~{\rm GeV}^{10}$ , the positivity of LHS
is also retained. But the minimum is still not wide enough, then the
finite threshold analysis should be performed. The results are shown
in the right part of Figure \ref{Mk}. At the Borel window, the mass
of $\kappa$ is close to 0.8 GeV.
To find the best sum rule, following the previous criteria, we find
that to $\kappa$, the convergence limits $M_B^2>0.25\rm {GeV}^2$ for
$s_0=0.8,~ 0.9 \rm {GeV}^2$ and $M_B^2>0.225,~0.3\rm {GeV}^2$ for
$s_0=0.7,~1.2 \rm {GeV}^2$, respectively. For instance, to $s_0=0.9
\rm {GeV}^2$, the convergence of OPE series is shown in Fig.
\ref{kc}.

The pole contributions for several values of threshold are listed in
Table II.
\begin{table}[hbt]
\caption{Pole contributions of various threshold.}
\begin{center}
\begin{tabular}{c|c|c|c|c}
 \hline  $s_{0}~({\rm GeV}^{2})$ & 0.7 & 0.8 & 0.9 & 1.2\\
 \hline   $M_B^{2}~({\rm GeV}^{2})$ & 0.225 & 0.25  & 0.25 & 0.5\\
 \hline   Pole (\%) & 43 &47 & 56 & 27\\
 \hline    $M_{\kappa}$ (GeV) & 0.75 & 0.8 & 0.82 & 0.95\\
  \hline
\end{tabular}
\end{center}
\end{table}
When $s_0=0.9{\rm GeV}^2$, $M_B^2=0.25{\rm GeV}^2$, we get a pole
contribution 56\%. Such a large pole contribution suggests that a
good sum rule has been obtained.  We get the mass of $\kappa$,
\begin{equation}\label{}
    m_{\kappa}=(820\pm80)\rm{MeV},~~~~~~~with~Pole~contribution(56\%).
\end{equation}
This pole contribution is also larger than that given by
\cite{zhu2}, where the pole contribution approaches 45\%.

\begin{figure}[hbt]
\begin{center}
\vspace{0ex}
\includegraphics[width=.42\textwidth]{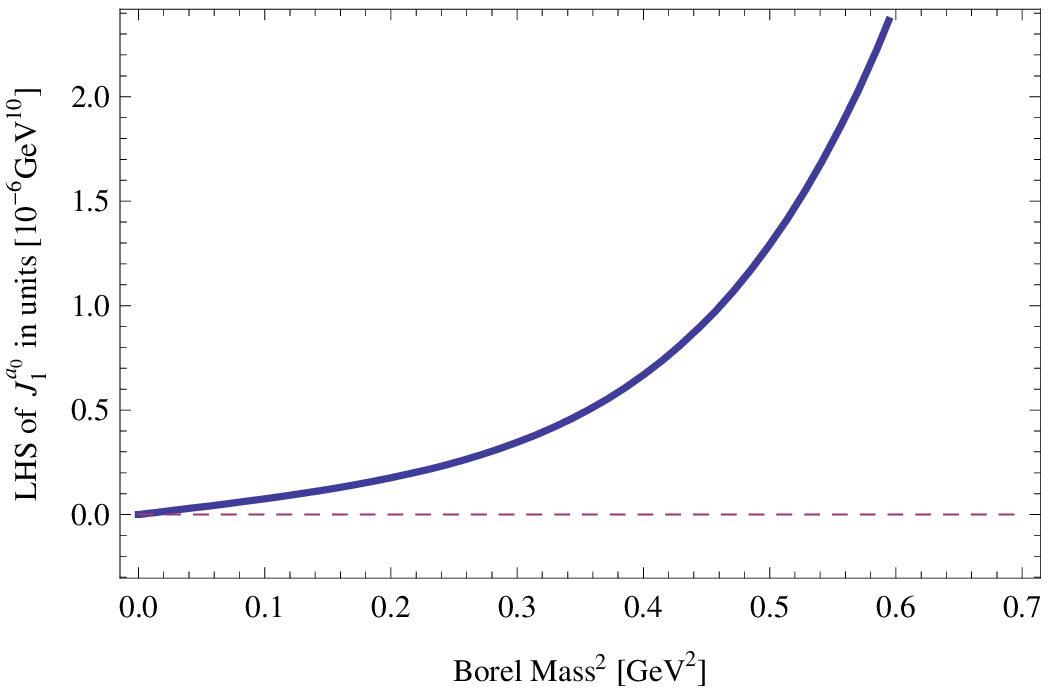}
\includegraphics[width=.42\textwidth]{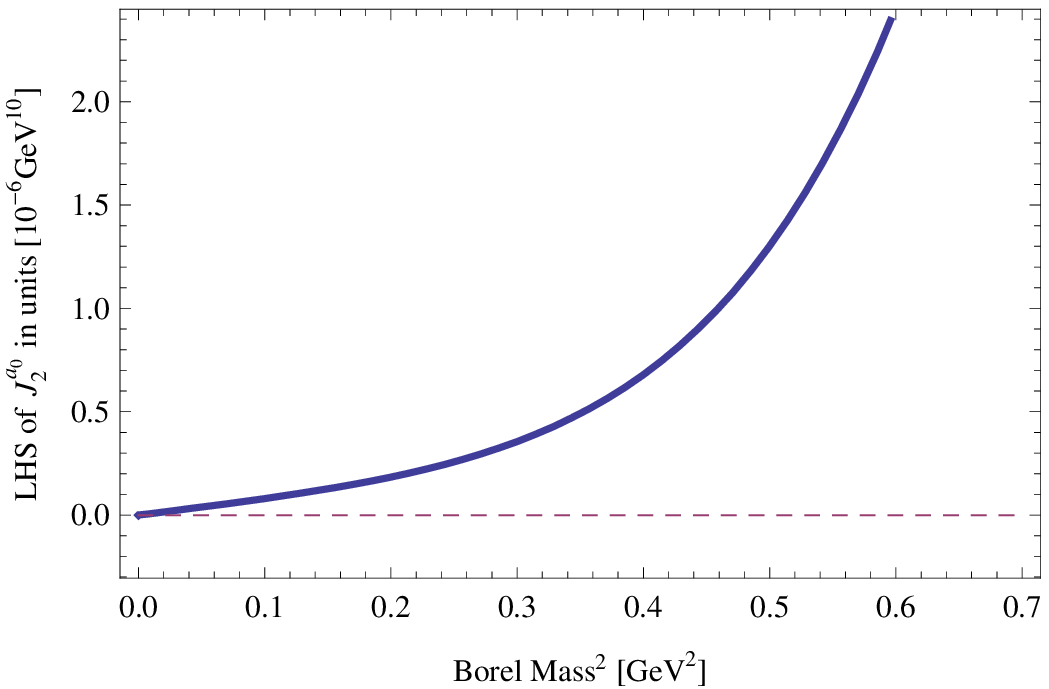}
\includegraphics[width=.42\textwidth]{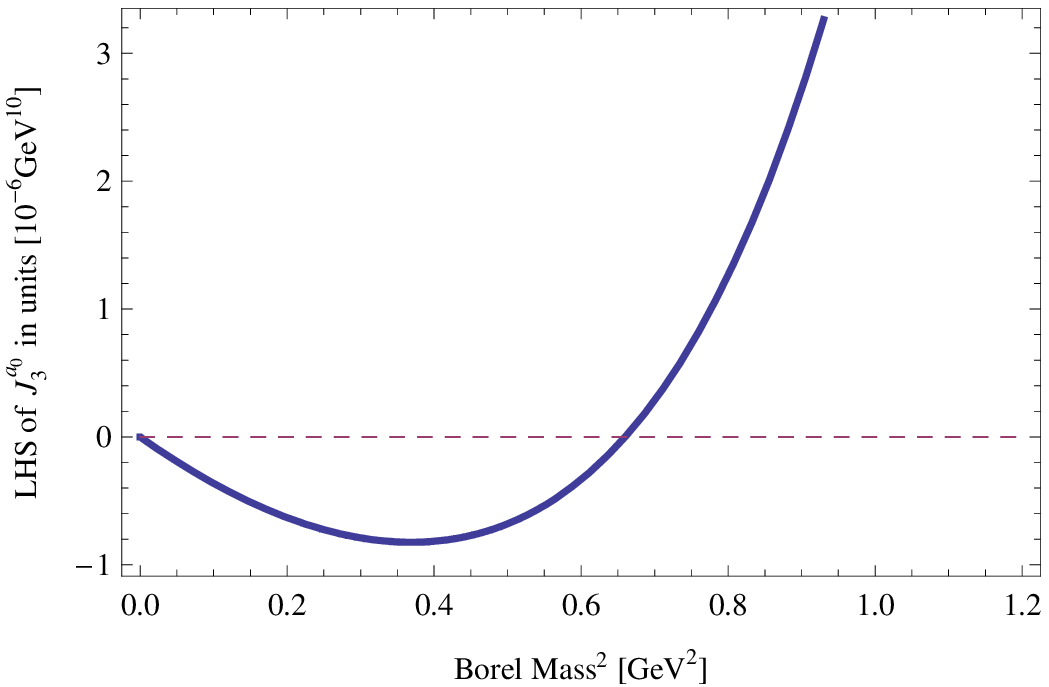}
\includegraphics[width=.42\textwidth]{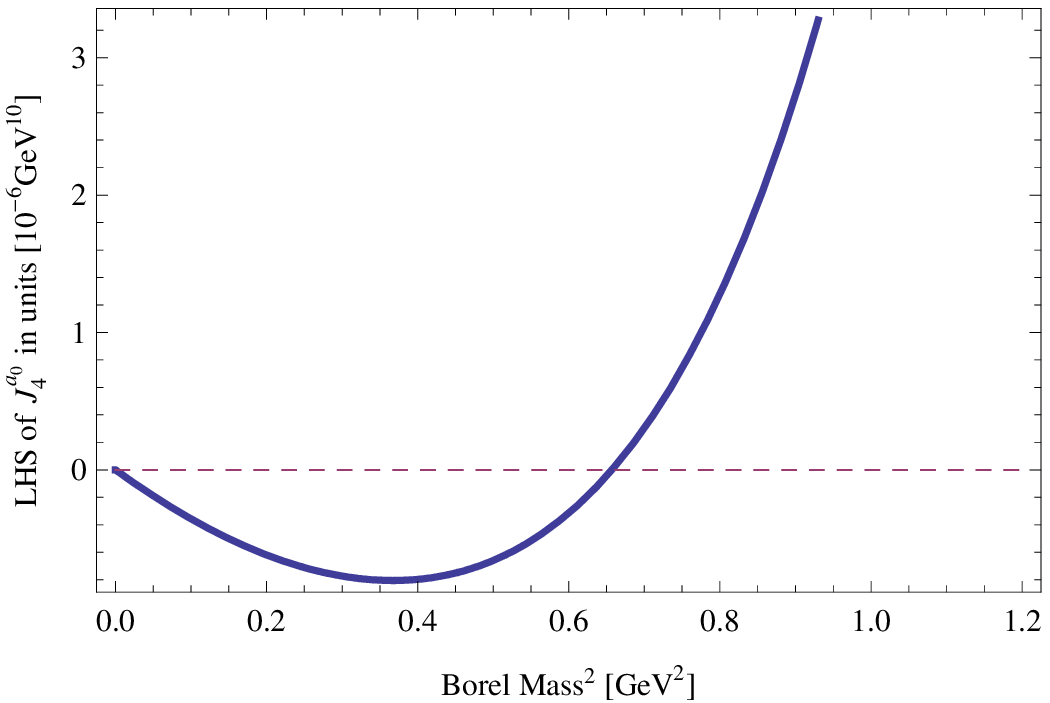}
\end{center}
\caption{\label{a1}LHS of four interpolating currents of $a_{+}$ and
$f_{0}$ as functions of Borel mass squared with $s_{0}$=infinity in
units of $\rm GeV^{10}$. }
\end{figure}

\begin{figure}[ht]
\begin{center}
\vspace{0ex}
\includegraphics[width=.43\textwidth]{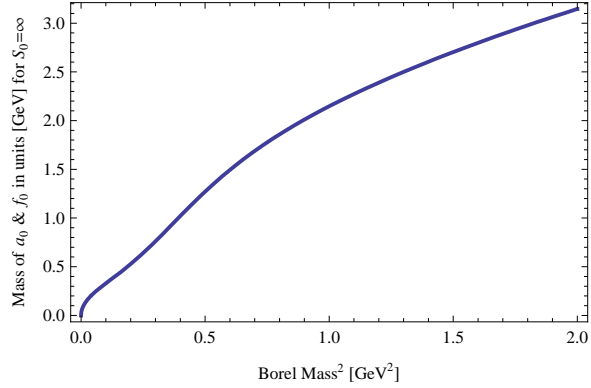}
\end{center}
\caption{\label{Ma}Mass of $a_{+}$ and $f_{0}$ as function of Borel
mass $M_{B}$ with $s_{0}$ being infinity. }
\end{figure}

Lastly, for $a_{+}$ and $f_{0}$ that are degenerate in OPE
calculations, the LHS of four possible interpolating currents are
shown in Fig. \ref{a1}, with threshold value $s_{0}$ being infinity.
Their numerical expressions are the following ones:
\begin{eqnarray}\label{spectrum-a0}
  \nonumber  \Pi^{a+,f0(\rm all)}_{1}&=& 1.9\times10^{-5}\M^{10}-2.3\times10^{-6}\M^{8}+4.0\times10^{-6}\M^{6}-5.8\times10^{-8}\M^{4}+7.2\times10^{-7}\M^{2} ,\\
  \nonumber \Pi^{a+,f0(\rm all)}_{2}&=&1.9\times10^{-5}\M^{10}-2.3\times10^{-6}\M^{8}+4.0\times10^{-6}\M^{6}-1.1\times10^{-8}\M^{4}+7.7\times10^{-7}\M^{2},  \\
 \nonumber  \Pi^{a+,f0(\rm all)}_{3}&=& 3.2\times10^{-6}\M^{10}-3.7\times10^{-7}\M^{8}+1.8\times10^{-6}\M^{6}+4.2\times10^{-6}\M^{4}-4.1\times10^{-6}\M^{2}, \\
 \Pi^{a+,f0(\rm all)}_{4}&=&3.2\times10^{-6}\M^{10}-3.7\times10^{-7}\M^{8}+1.8\times10^{-6}\M^{6}+4.2\times10^{-6}\M^{4}-4.0\times10^{-6}\M^{2}.
\end{eqnarray}
From Fig. \ref{a1} and above expressions, current $J_2^{a_+}$ seems
to be the best one. But when applying the traditional sum rule
method to estimate mass, it turns out that there is no minimum as
shown in Fig. \ref{Ma}. Furthermore, if we choose certain threshold
and Borel mass to reproduce the experimental center value of the
masses of $a_{+}$ and $f_{0}$, the pole contribution can only be
around 10\%. This indicates that in contrast to the success of SR
analysis of $\sigma$ and $\kappa$, the SR fails to analyze $a_+$ and
$f_0$, in terms of the interpolating currents deduced from their
wavefunctions as tetraquarks. The reason is as follows. Jaffe's
wavefunctions are the eigenfunctions of $H_{eff}$ in Eq. (\ref{H}).
However, $H_{eff}$ is only an approximate description of
color-magnetic interactions $H_{CM}=-\sum_{i\; j} C_{ij}
(\lambda_i\cdot\lambda_j)
(\overrightarrow{\sigma}_i\cdot\overrightarrow{\sigma}_j)$
\cite{DGG, hogaasen1, hogaasen2, dy}. If the flavor
$SU(3)_f$-symmetry is exact, the interaction strengthes $C_{ij}$ are
flavor-$(ij)$ independent, i.e., $C_{ij}=C$, then $H_{CM}=H_{eff}$.
But for real QCD, the constituent mass $m^{c}_u\approx m^{c}_d$,
while $m^{c}_s>\hat{m^{c}}\equiv(m^{c}_{u}+m^{c}_d)/2$. So $SU(3)_f$
must be broken within order
$\mathcal{O}((m^{c}_s-\hat{m^{c}})/m^{c}_s)\sim \mathcal{O}(0.3)$.
Therefore, both $H_{eff}$ and Jaffe's wavefunction
$|0^+,\underline{9} \rangle $ will suffer of this $SU(3)_f$ breaking
effect. In other words, $|0^+,\underline{9} \rangle $ can only be
thought of as the leading term of the eigenfunction of $H_{CM}$,
without considering the correction from the next leading term caused
by the strange quark content in $0^+$-tetraquarks. In
$\sigma(\{ud\}\{\bar{u}\bar{d}\})$, there is no strange quark, so no
such kind of corrections, hence $|\sigma\rangle=|0^+,\underline{9}
\rangle_{\sigma}$ is suitable. In
$\kappa(\{ud\}\{\bar{d}\bar{s}\})$, there is one strange quark, its
correction is relatively small, and the wavefunction
$|0^+,\underline{9} \rangle_{\kappa}$ may be still valid to some
extent. This is supported by numerical results. However, for
$f_0({1\over
\sqrt{2}}(\{us\}\{\bar{u}\bar{s}\}+\{ds\}\{\bar{d}\bar{s}\}))$ or
$a_+(\{us\}\{\bar{d}\bar{s}\})$, there are two strange quarks, the
$SU(3)_f$ breaking effects is doubled. To these cases, one cannot
insist the Jaffe's wavefunctions $|f_0\rangle=|0^+,\underline{9}
\rangle_{f_0}$ and $|a_+\rangle=|0^+,\underline{9} \rangle_{a_+}$ be
still good enough to describe the non-perturbative QCD physics.
Above all, we speculate that a legitimate SR analysis for $f_0$ and
$a_+$ should be based on the tetraquark's color-magnetic
wavefunctions which are more precise, encoding the
$SU(3)_f$-symmetry breaking effects.

\section{The direct instanton contribution to sum rule}
\subsection{Analytic results }
\begin{figure}[ht]
\begin{center}
\vspace{0ex}
\includegraphics[width=.43\textwidth]{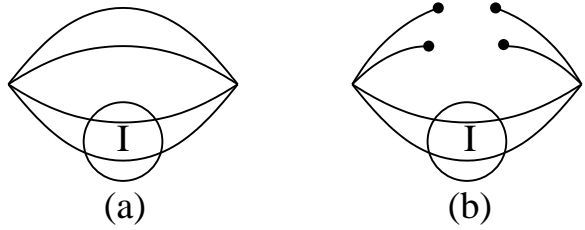}
\end{center}
\caption{\label{InstantiInst}The leading direct instanton
contribution to the correlator, where ``I'' represents the
instanton.}
\end{figure}

In addition to the contribution of power type from the OPE expansion
to the QCD SR, there are exponential contributions coming from
direct instanton contributions. The direct instantion contributions
originate from 't Hooft's instanton induced interaction
\cite{tHooft}. If the physics considered is relevant to two flavors,
instanton effects induce a four-fermion interaction, as illustrated
in Fig. \ref{InstantiInst} (usually called two-body single instanton
contribution defined in \cite{lee2}). In the framework of sum rule,
this kind of instanton effect can be encoded in the quark
propagator. Now the quark propagator has two terms,
\begin{equation}\label{}
    S^{q}_{ab}=S^{q({\rm st})}_{ab}+S^{q({\rm inst})}_{ab}.
\end{equation}
$S^{q({\rm st})}_{ab}$ corresponds to standard quark propagator
(Eqs. (\ref{propagator1}) and (\ref{propagator2})) in Euclidean
space, $S^{q({\rm inst})}_{ab}$ is related to instanton contribution
and can be calculated by using the following formula in Euclidean
space and regular gauge,
\begin{equation}\label{inst-propagator}
    S^{q({\rm inst})}_{ab}=A_{q}(x,y)\gamma_{\mu}\gamma_{\nu}(1+\gamma_{5})(U\tau_{\mu}^{+}\tau_{\nu}^{-}U^{\dagger})_{ab},
\end{equation}
where
\begin{equation}\label{}
    A_{q}(x,y)=-i\frac{r^{2}}{16\pi^{2}m_{q}^{\ast}}\phi(x-z_{0})\phi(y-z_{0})
\end{equation}
and
\begin{equation}\label{}
   \phi(x-z_{0})=\frac{1}{[(x-z_{0})^{2}+r^{2}]^{3/2}}.
\end{equation}
Here $r$ stands for the instanton size, $z_{0}$ for the center of
the instanton. $U$ represents the color orientation matrix of the
instanton in $SU(3)_{c}$ and $\tau^{+,-}_{\mu,\nu}$ are $SU(2)_{c}$
matrices. The effective mass of quark on the instanton vacuum is
$m_{q}^{\ast}=m_{q}-2\pi^{2}r^{2}_{c}\q2/3$ with current quark mass
$m_{q}$, here $q\in\{u, d, s\}$. At the final stage, we multiply the
result by a factor of two to take into account the anti-instanton
effect and integrate over the color orientation and instanton size.
When integrating over the instanton size, Shuryak's instanton liquid
model \cite{schafer} for QCD vacuum with density
$n_{r}=n_{eff}\delta(r-r_{c})$ has been used.

With the definition $Q^{2}=-q^{2}$, the direct instanton
contributions to the scalar nonet are listed below, corresponding to
above two diagrams. Here, we only exhibit the contributions to
$\sigma$-correlator, and the reader can find the results of other
tetraquarks in appendix. We denote the total contributions from
intanton and anti-instanton by ``inst''. Recalling that the direct
instanton contribution is possible only for different quark flavors,
so in case of $\sigma$, there is no direct three-body instanton
contribution (from instanton induced six-fermion interaction). But
to $\kappa$, $a_+$, $f_0$, three-body instanton contribution might
be important. However, in this paper, we only present the two-body
instanton contributions for these mesons, to capture the main
physics.
\begin{eqnarray}\label{inst-sigma}
    \Pi_{TT}^{\sigma(\rm inst)}&=&\frac{156n_{eff}r_{c}^{4}\q2^{2}}{3\pi^{4}m_{q}^{\ast2}}f_{0}(Q),\\
    \Pi_{SS}^{\sigma(\rm inst)}&=&\frac{32n_{eff}r_{c}^{4}}{\pi^{8}m_{q}^{\ast2}}f_{6}(Q)+\frac{19n_{eff}r_{c}^{4}\q2^{2}}{18\pi^{4}m_{q}^{\ast2}}f_{0}(Q),\\
    \Pi_{PP}^{\sigma(\rm
    inst)}&=&-\frac{32n_{eff}r_{c}^{4}}{\pi^{8}m_{q}^{\ast2}}f_{6}(Q)+\frac{19n_{eff}r_{c}^{4}\q2^{2}}{18\pi^{4}m_{q}^{\ast2}}f_{0}(Q),\\
    \Pi_{TS}^{\sigma(\rm inst)}&=&\Pi_{ST}^{\sigma(\rm
    inst)}=\frac{2n_{eff}r_{c}^{4}\q2^{2}}{\pi^{4}m_{q}^{\ast2}}f_{0}(Q),\\
    \Pi_{TP}^{\sigma(\rm inst)}&=&\Pi_{PT}^{\sigma(\rm inst)}=\frac{2n_{eff}r_{c}^{4}\q2^{2}}{\pi^{4}m_{q}^{\ast2}}f_{0}(Q),\\
    \Pi_{AA}^{\sigma(\rm inst)}&=&\frac{48n_{eff}r_{c}^{4}}{\pi^{8}m_{q}^{\ast2}}f_{6}(Q)+\frac{68n_{eff}r_{c}^{4}\q2^{2}}{9\pi^{4}m_{q}^{\ast2}}f_{0}(Q),\\
    \Pi_{AS}^{\sigma(\rm inst)}&=&\Pi_{SA}^{\sigma(\rm inst)}=-\frac{20n_{eff}r_{c}^{4}}{\pi^{8}m_{q}^{\ast2}}f_{6}(Q),\\
    \Pi_{AP}^{\sigma(\rm inst)}&=&\Pi_{PA}^{\sigma(\rm inst)}=-\frac{20n_{eff}r_{c}^{4}}{\pi^{8}m_{q}^{\ast2}}f_{6}(Q).
\end{eqnarray}
In above expressions,
\begin{eqnarray}
 \nonumber f_{6}(Q) &=& \int d^{4}z_{0}\int d^{4}x\frac{e^{iq\cdot x}}{x^{6}[z_{0}^{2}+r_{c}^{2}]^{3}[(x-z_{0})^{2}+r_{c}^{2}]^{3}}, \\
  f_{0}(Q) &=& \int d^{4}z_{0}\int d^{4}x\frac{e^{iq\cdot x}}{[z_{0}^{2}+r_{c}^{2}]^{3}[(x-z_{0})^{2}+r_{c}^{2}]^{3}}.
\end{eqnarray}
The Borel transformation of $f_{6}(Q)$ and $f_{0}(Q)$ are:
\begin{eqnarray}\label{Borel-f}
\nonumber  \hat{B}[f_{6}(Q)] &=& -\frac{\pi^{4}M^{12}_{B}}{2^{13}}\int^{1}_{0}dt \int^{1}_{0}dy\frac{e^{-M^{2}_{B}r_{c}^{2}/(4ty(1-y))}}{y^{2}(1-y)^{2}}(X^{2}+5X^{3}+10X^{4}\\
\nonumber   &&+10X^{5}+5X^{6}+X^{7}),\\
    \hat{B}[f_{0}(Q)]&=&\frac{\pi^{4}M^{6}_{B}}{16}e^{-M^{2}_{B}r_{c}^{2}/2}(K_{0}(M^{2}_{B}r_{c}^{2}/2)+K_{1}(M^{2}_{B}r_{c}^{2}/2)),
\end{eqnarray}
where we adopt the notations in paper \cite{lee2}, $X=(1-t)/t$ and
$K_{n}(x)$ is the McDonald function.
\subsection{Numeric analysis of QCD sum rule with instanton effects}
To evaluate the direct instanton effects quantitatively, we make use
of the following relation between the parameters of Shuryak
instanton model \cite{schafer}.
\begin{equation}\label{}
    \frac{n_{eff}}{m_{q}^{\ast2}}=\frac{3}{4\pi^{2}r_{c}^{2}}~~~~~q\in\{u,~d\},
\end{equation}
with
\begin{equation}\label{}
    r_{c}=1.6~\mbox{GeV}^{-1}.
\end{equation}
Considering the single instanton effects, the left hand sum rule
becomes:
\begin{equation}\label{}
    \Pi_{\rm LHS}(Q^{2})=\Pi^{\rm OPE}(Q^{2})+\Pi^{\rm inst}(Q^{2}).
\end{equation}
After Borel transforming the both side of the QCD sum rule, we
obtain the following relation
\begin{equation}\label{2-contribution}
  \B_{M_{B}^{2}}\Pi^{\rm OPE}(Q^{2})+\B_{M_{B}^{2}}\Pi^{\rm inst}(Q^{2})=2\pi f_{X}^{2}m_{X}^{8}e^{-m_{X}^{2}/M_{B}^{2}}.
\end{equation}
In above expressions,
\begin{equation}\label{}
\B_{M_{B}^{2}}\Pi^{{\rm OPE}}(Q^{2})=\int^{S_{0}}_{0}
    e^{-s/M_{B}^{2}}\rho^{\rm OPE}(s)ds,
\end{equation}
where we have chosen a finite threshold to suppress the contribution
from continuum. Utilizing the results in previous sections, the left
hand sum rule can be performed for each possible interpolating
current in (\ref{current4}) belonging to a certain meson. Then we
can make use of the best current to fit the right hand sum rule to
obtain the mass and residue. This approach was first suggested by
\cite{lee2}. In the following, for the sake of simplicity, we will
only present a detailed analysis for $\sigma$ meson. For other
mesons,  the results are also exhibited.

\begin{figure}[hbt]
\begin{center}
\vspace{0ex}
\includegraphics[width=.4\textwidth]{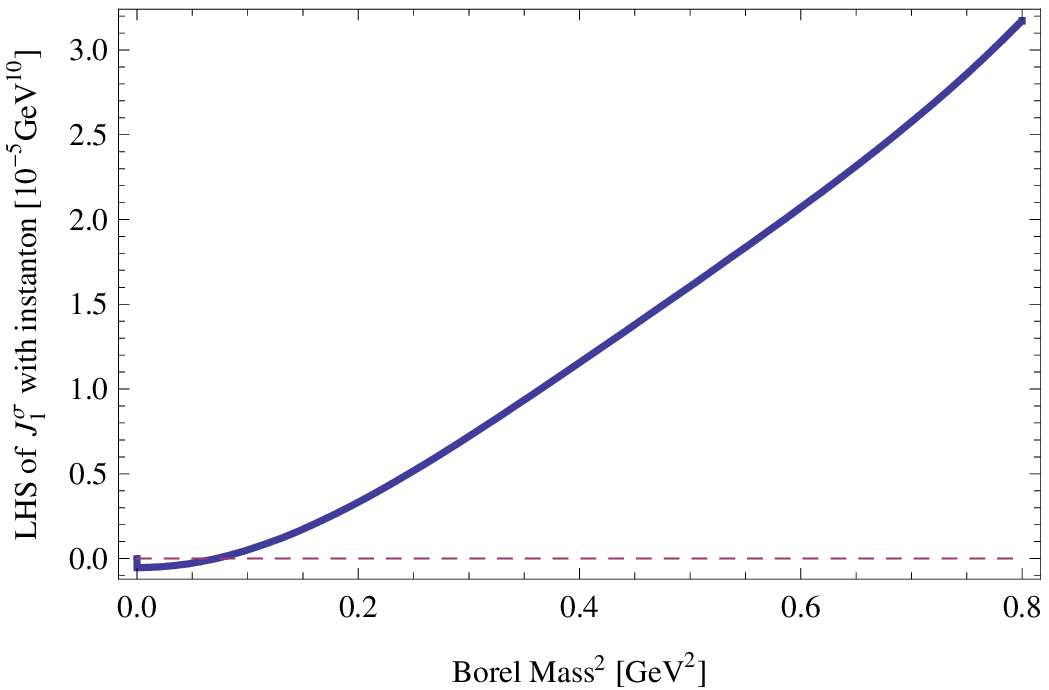}
\includegraphics[width=.4\textwidth]{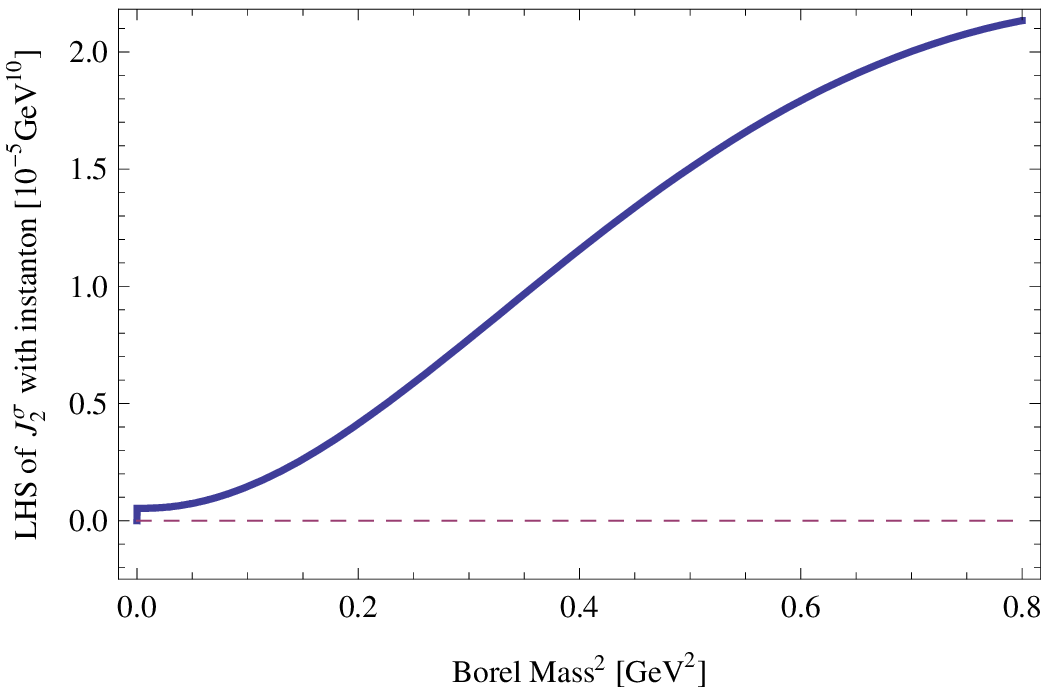}
\includegraphics[width=.4\textwidth]{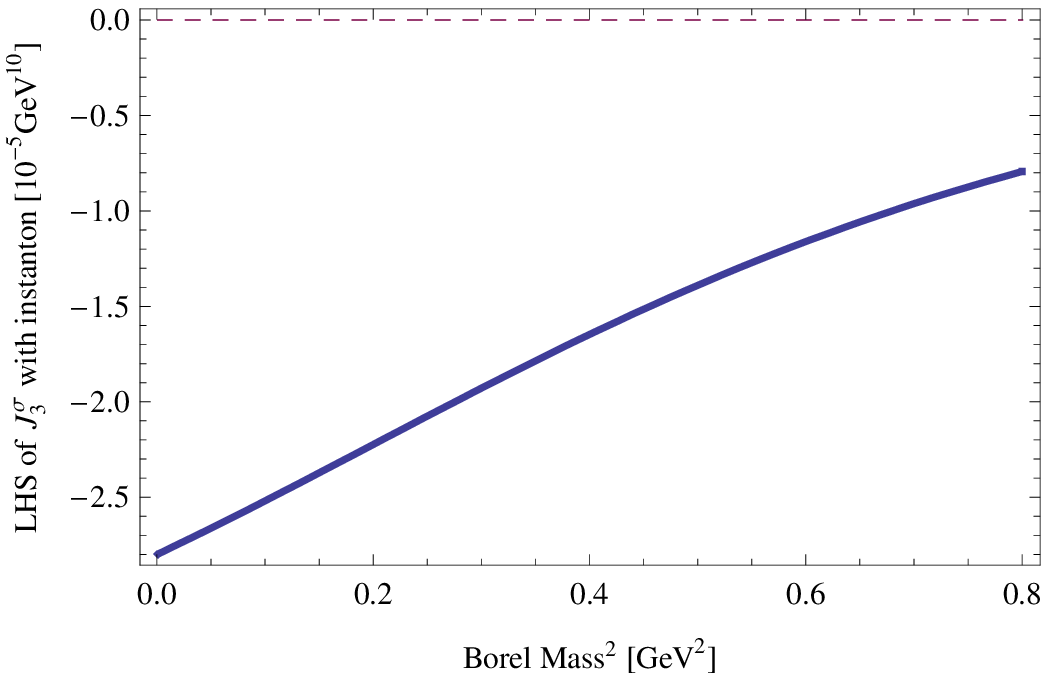}
\includegraphics[width=.4\textwidth]{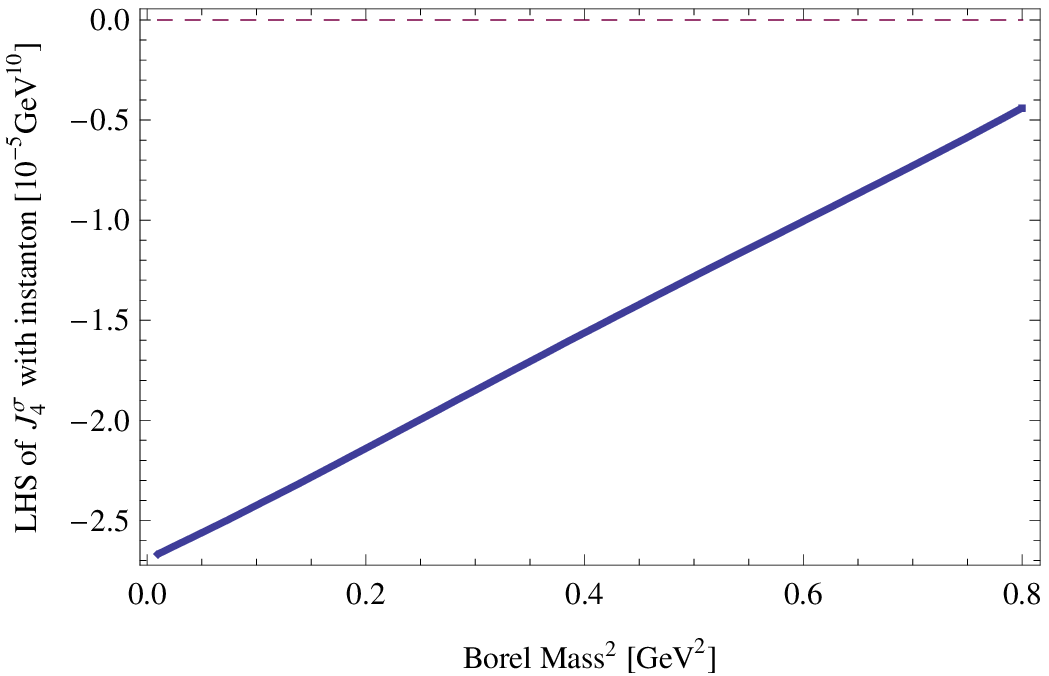}
\end{center}
\caption{\label{li}LHS of $\sigma$ including the instanton effects
of four interpolating currents with $s_{0}$=0.6 $\mbox{GeV}^2$. }
\end{figure}

 In Fig. \ref{li}, we show the Borel transformed correlators
$\Pi(M_{B}^{2})$, including the instanton effects, at threshold
value $s_{0}$=0.6 $\mbox{GeV}^2$. From the Figure, we see that the
instanton contributions are not always positive. To current
$J_1^{\sigma}$, they provide little negative contributions, and
spoil the positivity of LHS obviously, when Borel mass is small; to
current $J_3^{\sigma}$ and $J_4^{\sigma}$, instanton effects make
the LHS rather negative, and this may be the usually called
dangerous instanton contribution to sum rule \cite{lee2}; only to
current $J_2^{\sigma}$, the instanton effects improve the OPE
calculation completely. This feature can be seen more clearly, if we
notice that in Eqs. (\ref{inst-sigma})-(79):
\begin{eqnarray}
\nonumber  \Pi_{TT}^{\sigma(\rm inst)}&=&\frac{156n_{eff}r_{c}^{4}\q2^{2}}{3\pi^{4}m_{q}^{\ast2}}f_{0}(Q),\\
\nonumber   \Pi_{PP}^{\sigma(\rm inst)}&=&-\frac{32n_{eff}r_{c}^{4}}{\pi^{8}m_{q}^{\ast2}}f_{6}(Q)+\frac{19n_{eff}r_{c}^{4}\q2^{2}}{18\pi^{4}m_{q}^{\ast2}}f_{0}(Q),\\
    \Pi_{TP}^{\sigma(\rm inst)}&=&\Pi_{PT}^{\sigma(\rm inst)}=\frac{2n_{eff}r_{c}^{4}\q2^{2}}{\pi^{4}m_{q}^{\ast2}}f_{0}(Q).
\end{eqnarray}
In above expressions, the coefficients of $f_{0}(Q)$ are positive,
while the coefficient of $f_{6}(Q)$ is negative. After Borel
transformation, $f_{0}(Q)$ and $f_{6}(Q)$ are just as in Eq.
(\ref{Borel-f}). Numerically, $\hat{B}[f_{0}(Q)]$ is always
positive, but $\hat{B}[f_{6}(Q)]$ is always negative, so totally,
the instanton contributions to the current $J_2^{\sigma}$ are
positive. From Fig. \ref{li}, it is clear that the instanton
contributions improve the convergence of current $J_2^{\sigma}$ when
Borel mass is small.
At this moment, we can use the numeric results associated with LHS
of current $J_2^{\sigma}$, at threshold value 0.6 $\mbox{GeV}^2$, to
fit the RHS in single resonance approximation that is just the Eq.
(\ref{2-contribution}), as illustrated in Fig. \ref{LR}. That
choosing 0.6 $\mbox{GeV}^2$ as the value of threshold is inspired by
previous OPE results. The fitted mass and residues are listed in
Table 3:

\begin{figure}[ht]
\begin{center}
\vspace{0ex}
\includegraphics[width=.42\textwidth]{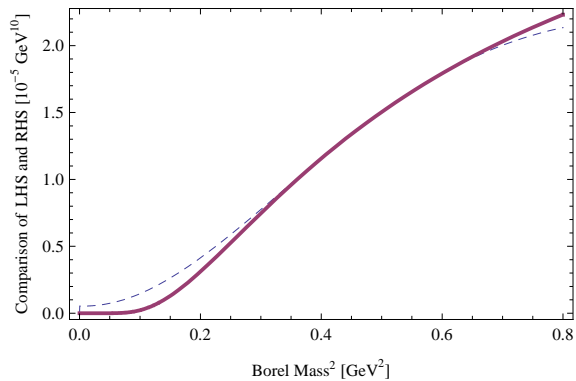}
\end{center}
\caption{\label{LR}The dashed and thick lines represent the left
hand sum rule and right hand sum rule, respectively. To RHS, the
mass and residue are presented in Table III.}
\end{figure}
From the table, we notice that after adding the instanton
contribution, the mass of $\sigma$ meson is still close to OPE
result Eq.(\ref{sigmaM}). Then the instanton contribution is
compatible with OPE results. It suggests that the physical mass of
$\sigma$ depends weakly on the choice of QCD vacuum.
\begin{table}[hbt]
\caption{Fitted masses and residues in single resonance
approximation}
\begin{center}
\begin{tabular}{c|c|c}
 \hline  $s_{0}({\rm GeV}^{2})$ & $M_{\sigma}$(GeV) & ${ f}_{\sigma}  (10^{-2}\rm GeV)$ \\
 \hline   0.6 & 0.72 & 0.94  \\
 \hline   1  & 0.73 &0.93 \\
  \hline
\end{tabular}
\end{center}
\end{table}

In the case of $s_0=0.6\rm{GeV^2}$, considering a $(\pm10\%)$
variation of instanton size $r_c=1.6\pm0.2$, we find a corresponding
variation of $m_{\sigma}=720^{-100}_{+~60}\rm{MeV}$ and $
f_{\sigma}=0.94^{+~0.4}_{-0.07}(10^{-2}\rm GeV)$. It seems like that
the change of physical quantity lies within an acceptable range and
the residue is more sensitive to the variation of instanton size
compared with the mass. In \cite{Kojo}, the authors discussed the
meaning of the residue. In their notations, residue is defined as
$\lambda^2=2\pi f_X^2M_X^8$. So we obtain a residue
$\lambda^2=4\times10^{-5}{\rm GeV^{10}}$, which is larger than
$\lambda^2=2\times10^{-6}{\rm GeV^{10}}$ presented in \cite{Kojo}.
According to the explanation of \cite{Kojo}, large residue signifies
the interpolating current operators have enough overlaps to the
resonance states and the sum rule constructed with approximate OPE
may contain enough information for the resonance to be extracted. So
in our case, evaluating OPE up to dimension eight condensates seems
reasonable.

Finally, in order to investigate further the widths of the $\sigma$
meson states, it is necessary to find out three point correlation
functions for $\sigma\rightarrow\pi\pi$, which has got out of the
scope of this paper.

As for other mesons, the current $J_2$ still shows the best
performance. The fitted masses and residues for $\kappa$, $a_{+}$
and $f_{0}$ are presented in Table IV, V and VI in appendix ,
respectively.

\section {Conclusion and Discussion } In this paper,
 we study the $0^{+}$ nonet mesons as tetraquark
 states
 with interpolating currents induced from the color-magnetic
 wavefunction. This wavefunction is the eigenfunction of the effective color-magnetic Hamiltonian with the
 lowest eigenvalue, meaning that the state with this wavefunction is the most stable one and is most probable to be observed in
 experiments.
 Our approach can be recognized as
 constructing interpolating currents dynamically.
 We find that based on a current which is a proper mixture of the
 tensor and pseudoscalar contents, a good sum rule can be obtained. Our result can be perceived as a direct support to multiquark scenario described by the
 color-magnetic interaction, by means of QCD sum rule.

In the SR calculations performed in this paper, we have taken into
account the contributions from operators up to dimension $d=8$ in
the OPE. The results of
 SR analysis without instanton effects for $0^+$ meson nonet
$\{\sigma,\;\kappa,\;f_0,\;a_+\}$ are :
\begin{enumerate}
\item  $\sigma$:
  In the SR analysis ,  a good Borel stability turns out in the region
$M_B^2>0.2~{\rm GeV}^2$. Taking $M_B^2\approx 0.2~{\rm GeV}^2$ and
the threshold $s_0\approx0.6~{\rm GeV}^2$, the largest pole
contribution is $52\%$ implying that a good SR analysis is achieved.
Where we extract the mass of $\sigma$ $(600\pm75)$ MeV.

\item $\kappa$: A good sum rule was found when $s_0=0.9\rm{GeV}^2$,
$M^2_B>0.25\rm{GeV}^2$. We obtain $\kappa$ mass $(820\pm80)$MeV with
pole contribution approaching 56\%.

\item $f_{0}~{\rm and}~a_{+}$: to obtain a mass about 1 GeV by choosing the threshold and Borel mass,
 the pole contributions in SR are always around 10\%. This
indicates that the SR fails to analyze $a_+$ and $f_0$ by using the
interpolating currents deduced from the wavefunctions. We guess the
reason is that in $f_0({1\over
\sqrt{2}}(\{us\}\{\bar{u}\bar{s}\}+\{ds\}\{\bar{d}\bar{s}\}))$ or
$a_+(\{us\}\{\bar{d}\bar{s}\})$, there are two strange quarks, so
$SU(3)_f$ breaking effects are too strong to be negligible. This
causes the Jaffe's wavefunctions $|f_0\rangle=|0^+,\underline{9}
\rangle_{f_0}$ and $|a_+\rangle=|0^+,\underline{9} \rangle_{a_+}$ to
miss some aspects of the $f_0$- and $a_+$-physics. We speculate that
a legitimate SR analysis for them should be based on the tetraquark
color-magnetic wavefunctions including the $SU(3)_f$-breaking
effects due to $m_s^c>\hat{m}^c$.
\end{enumerate}

 Proceed stepwise, we consider the direct instanton contribution.
 To the current $J_2$, the instanton effects are completely positive.
 Numerically, this positive effects improve the
 small Borel mass behavior of the Borel transformed correlator of current $J_2$. Meanwhile, adding instanton
 effects, the LHS gives a result compatible with OPE results.

 Finally, we go one step further and believe
 that the idea demonstrated in this paper also applies to
 $0^-$-$q^3\bar{q}^3$ system. In \cite{DPY},
 the authors have successfully
  extended Jaffe's method from $q^2\bar{q}^2$ to $q^3\bar{q}^3$
six-quark system (i.e., baryonium). One of the non-trivial results
in \cite{DPY} for baryonium is the existance of a counterpart of
$\sigma$. We denote this state by $|0^-, \underline{1}_f\rangle$.
Corresponding  to Eq. (\ref{5}) for tetraquark, \cite{DPY} shows
\begin{equation}\label{100}
H_{eff}|0^-, \underline{1}_f\rangle =-82.533\widetilde{C} |0^-,
\underline{1}_f\rangle.
\end{equation}
In baryonium contents, its color-spin-flavor wavefunction can be
expressed as:
\begin{eqnarray}
\nonumber|0^-, \underline{1}_f\rangle &\equiv&|\mathbf{1},\mathbf{}1_f\otimes\mathbf{1}_f\rangle_1=0.591|(\mathbf{56}_{cs},\mathbf{10}_c,\mathbf{4};\mathbf{1}_f),(\overline{\mathbf{56}}_{cs},\overline{\mathbf{10}}_c,\mathbf{4};\mathbf{1}_f),\mathbf{1}_c,\mathbf{1};\mathbf{1}_f\otimes\mathbf{1}_f\rangle\\
\label{101}&&+0.807|(\mathbf{56}_{cs},\mathbf{8}_c,\mathbf{2};\mathbf{1}_f),(\overline{\mathbf{56}}_{cs},\mathbf{8}_c,\mathbf{2};\mathbf{1}_f),\mathbf{1}_c,\mathbf{1};\mathbf{1}_f\otimes\mathbf{1}_f\rangle,
\end{eqnarray}
where the notations in \cite{DPY} have been used. Like
$|\sigma\rangle$, $|0^-, \underline{1}_f\rangle$ has the largest
mass defect among all the baryoniums. This implies that $|0^-,
\underline{1}_f\rangle$, the lightest baryonium meson, may represent
 a stable physical state. Like Eq. (\ref{8}), the mass of $|0^-,
\underline{1}_f\rangle$ can be estimated roughly in the naive
constituent quark model as follows
\begin{eqnarray}
\nonumber m_{|0^-, \underline{1}_f\rangle}&\approx& \langle \sum_i
m_i^c-\widetilde{C}\sum_{i\; j} (\lambda_i\cdot\lambda_j)
(\overrightarrow{\sigma}_i\cdot\overrightarrow{\sigma}_j)\rangle_{|0^-,
\underline{1}_f\rangle}\\
\nonumber &\approx& (4\times 360{\rm MeV}+2\times 540{\rm MeV})
-82.533\times \left ( {4\times 20{\rm MeV} + 2\times 15 {\rm
MeV}\over 6} \right )\\
\label{102} &\approx& 1.007{\rm GeV}.
\end{eqnarray}
We find that the mass of $|0^-, \underline{1}_f\rangle$ is close to
that of $\eta'(960)$ \cite{Yao:2006px}. Furthermore, their quantum
numbers are the same. So in the multiquark picture, we might
identify $|0^-, \underline{1}_f\rangle$ as $\eta'(960)$, or perceive
$\eta'(960)$ as a baryonium or a Fermi-Yang meson \cite{FY}.
Alternatively, there may be a large weight baryonium component in
$\eta'(960)$. Usually, in the $q\bar{q}$-picture, the mass of
$\eta'$ is attributed to $U(1)_A$ anomaly with non-trivial $\theta $
vacuum in QCD \cite{tHooft}. However, that scenario has not excluded
other schemes yet (e.g., see \cite{donoghue}). In our case, a
further examination to the conjecture on $\eta'$ in non-perturbative
QCD should be meaningful. Since we have already known the
color-magnetic wavefunction for $|0^-, \underline{1}_f\rangle$,
following the method presented in this paper, a SR analysis is
straightforward. The result will be helpful to understand two
interesting experimental measurements that may reveal the baryonium
content of $\eta'$. Those experiments are that:

i) to measure the anomalous enhancement near the mass threshold in
the $p\bar{p}$ invariant-mass spectrum from $J/\psi\rightarrow
\gamma p\bar{p}$ reported by BES \cite{BES1}.

ii) to observe resonance $X(1835)$ in $J/\psi\rightarrow \gamma
\pi^+ \pi^- \eta'$ \cite{BES2}. In \cite{BES1} the data fitting
indicates that the enhancement is a S-wave Breit-Wigner resonance
 $X(1835)$ \cite{1}. It has been estimated that the decay branching fraction
 $B(X\rightarrow p\bar{p})>4\%$ \cite{2}. The decay mode of
 $X\rightarrow p\bar{p}$ is due to the tail effect of
 enhancement resonance of $X(1835)$ near the threshold of process
 $J/\psi\rightarrow \gamma p\bar{p}$, therefore the fact of $B(X\rightarrow p\bar{p})>4\%$
 means the coupling between $X$ and $p\bar{p}$ is very very strong.
The most natural interpretation to this fact is that $X(1835)$ is
simply a bound state of $p-\bar{p}$. Namely, $X(1835)$ is a
$q^3\bar{q}^3$-baryonium molecular state \cite{datta, yan}. In
another hand, the major decay mode for $X(1835)$ is
$X(1835)\rightarrow \pi^+\pi^-\eta'$ observed by BES \cite{BES2}. It
indicates that $X(1835)$ is a molecular exciting state of meson
$\eta'$ \cite{yan}. Consequently, the quark component  of $\eta'$
should be same as $X(1835)$, i.e., $\eta'$ would be a
$0^-$-baryonium meson, or a meson with large weight baryonium
component. BES observations \cite{BES1, BES2} provide evidence to
this multiquark picture for $\eta'$ meson.

\section* {ACKNOWLEDGEMENTS}
\indent We would like to thank R. L. Jaffe for helpful comments to
this work and information discussions. We also thank Gui-Jun Ding,
Dao-Neng Gao, N. I. Kochelev, Jia-Lun Ping for discussions and Yi
Wang, Tower Wang for warm helps. Especially, we are grateful to
Shi-Lin Zhu for introducing useful OPE calculation method to us.
This work is partially supported by National Natural Science
Foundation of China under Grant Numbers 90403021, and by  the
Chinese Science Academy Foundation under Grant Numbers KJCX-YW-N29.

\begin{appendix}
\section{}
\subsection{Formulas of necessary spectral functions of $\kappa$, $a_+$ and
$f_0$} For $\kappa$ $(\{ud\}\{\bar{s}\bar{d}\})$, since the current
mass $m_{s}$ is much bigger than $m_{u}, m_{d}$, we can ignore terms
proportional to $m_{u}, m_{d}$ when listing the necessary spectral
functions. Having done this, the length of formulas will be
shortened, and the reader can have a clear impression about the
structure of spectral functions. We will do the same thing for
$a_{+}$ and $f_{0}$. However, in numerical calculations, the
contributions from the $u, d$ quark mass terms have been taken into
account. The spectral functions are the followings:
\begin{eqnarray}\label{correusdd-TT}
\rho^{\kappa \rm OPE}_{T,T}&=&\frac{s^4}{1280 \pi^6} -\frac{m_s^2 }
{64 \pi^6}s^3 +( \frac{11\G2} {768 \pi^6} +\frac{m_s\s2}{8 \pi^4})
s^2 - \frac{11 m_s^2 \G2}{256\pi^6}s + \frac{11
m_s\G2\s2}{192\pi^4},\\
\nonumber&&\\
\nonumber\rho^{\kappa \rm OPE}_{S,S}&=&\frac{s^4} {61440 \pi^6}
-\frac{{m_s}^2 s^3} {3072 \pi^6} + ( \frac{\G2} {6144 \pi^6} -
\frac{{m_s}\q2} {192 \pi^4} + \frac{{m_s}\s2} {384 \pi^4}  ) s^2\\
\nonumber&& + ( - \frac{m_s^2 \G2}{2048\pi^6} - \frac{m_s \Gq
}{128\pi^4} + \frac{\q2^2}{24\pi^2} + \frac{\q2\s2 }{24\pi^2}) s
\\
 && - \frac{m_s^2 \q2^2}{12\pi^2} -
\frac{m_s\G2\q2}{768\pi^4} + \frac{m_s\G2\s2}{1536\pi^4} + \frac{\q2
\Gq}{24\pi^2} + \frac{\s2\Gq}{48\pi^2} + \frac{\q2\Gq}{48\pi^2},\\
\nonumber&&\\
\rho^{\kappa \rm OPE}_{T,S}&=&\rho^{\kappa \rm
OPE}_{S,T}=-\frac{\G2}{1024\pi^{6}}s^{2}+\frac{3\G2m_{s}^{2}}{1024\pi^{6}}s
 -\frac{\G2\s2m_{s}}{256\pi^{4}},\\
 \nonumber&&\\
 \nonumber\rho^{\kappa \rm OPE}_{P,P}&=&\frac{s^{4}}{61440\pi^{6}}-\frac{m_{s}^{2}}{3072\pi^{6}}s^{3}+(\frac{\q2m_{s}}{192\pi^{4}}+\frac{\s2m_{s}}{384\pi^{4}}+\frac{\G2}{6144\pi^{6}})s^{2}-(\frac{\q2^{2}}{24\pi^{2}}+\frac{\q2\s2}{24\pi^{2}}-\frac{\Gq m_{s}}{128\pi^{4}}\\
  &&+\frac{\G2m_{s}^{2}}{2048\pi^{6}})s+(\frac{m_{s}^{2}\q2^{2}}{12\pi^{2}}-\frac{\Gs\q2}{48\pi^{2}}-\frac{\Gq(\s2+2\q2)}{48\pi^{2}}+\frac{\G2(\q2+\s2)m_{s}}{1536\pi^{4}}),\\
 \nonumber&&\\
\rho^{\kappa \rm OPE}_{T,P}&=&\rho^{\kappa \rm OPE}_{P,T}
 =-\frac{\G2}{1024\pi^{6}}s^{2}+\frac{3\G2m_{s}^{2}}{1024\pi^{6}}s-\frac{\G2\s2m_{s}}{256\pi^{4}},\\
 \nonumber&&\\
\nonumber\rho^{\kappa \rm
OPE}_{A,A}&=&\frac{s^{4}}{7680\pi^{6}}-\frac{m_{s}^{2}}{384\pi^{6}}s^{3}+(\frac{m_{s}(\s2-\q2)}{48\pi^{4}}+\frac{5\G2}{3072\pi^{6}} )s^{2}-(\frac{\Gq m_{s}}{32\pi^{4}}+\frac{5\G2m_{s}^{2}}{1024\pi^{6}}-\frac{\q2\s2}{6\pi^{2}}\\
  &&-\frac{\q2\q2}{6\pi^{2}})s+(\frac{\Gq(2\q2+\s2)}{12\pi^{2}}-\frac{\q2^{2}m_{s}^{2}}{3\pi^{2}}+\frac{m_{s}\G2(5\s2-2\q2)}{768\pi^{4}}+\frac{\Gs\q2}{12\pi^{2}}),\\
\nonumber &&\\
\rho^{\kappa \rm OPE}_{A,S}&=&\rho^{\kappa\rm OPE}_{S,A}=\frac{\G2
m_{s}\q2}{256\pi^{4}},\\
\rho^{\kappa \rm OPE}_{A,P}&=&\rho^{\kappa \rm
OPE}_{P,A}=0.~~~~~~~~~~~~~~~~~~~~~~~~~~~~~~~~~~~~~~~~~~~~~~~~~~~~~~~~~~~~~~~~~~~~~~~~~~~~~~~~~
\end{eqnarray}

For $a_{+}~(\{us\}\{\bar{d}\bar{s}\})$ and
$f_{0}~(\frac{1}{\sqrt{2}}(\{su\}\{\bar{s}\bar{u}\}+\{sd\}\{\bar{s}\bar{d}\})$,
we only list the spectral functions for $a_{+}$ below. This is
because in the widely adopted scheme Eq. (\ref{propagator1}), $u$
and $d$ quark take
 the same value of current masses and condensates,
 which
 leads to
 a direct consequence that from the OPE calculation of the correlators of currents, we can not discern $a_{+}$ and
 $f_{0}$. In other words, to each kind
 interpolating current in Eq. (\ref{current4}), the correlators of $a_{+}$'s and the
 correlators of $f_{0}$'s take the same expressions after completing
 the OPE calculation.

\begin{eqnarray}\label{correuuss-TT}
 \nonumber  \rho^{a_{+}
 \rm OPE}_{T,T}&=&\frac{s^{4}}{1280}-\frac{m_{s}^{2}}{32\pi^{6}}s^{3}+(\frac{3m_{s}^{4}}{16\pi^{6}}+\frac{\s2m_{s}}{4\pi^{4}}+\frac{11\G2}{768})s^{2}-(\frac{3\s2 m_{s}^{3}}{2\pi^{4}}+\frac{11\G2m_{s}^{2}}{128\pi^{6}})s+(\frac{4\q2^{2}m_{s}^{2}}{\pi^{2}}\\
&&+\frac{\s2^{2}m_{s}^{2}}{\pi^{2}}+\frac{5\G2m_{s}^{4}}{128\pi^{6}}+\frac{11\G2\s2m_{s}}{96\pi^{4}}),\\
\nonumber&&\\
\label{correuuss-SS}
\nonumber  \rho^{a_{+} \rm OPE}_{S,S}&=& \frac{s^{4}}{61440\pi^{6}}-\frac{m_{s}^{2}}{1536\pi^{6}}s^{3}+(\frac{m_{s}^{4}}{256\pi^{6}}+\frac{m_{s}(\s2-2\q2)}{192\pi^{4}}+\frac{\G2}{6144\pi^{6}})s^{2}-(\frac{m_{s}^{3}\s2}{32\pi^{2}}-\frac{\q2\s2}{12\pi^{2}}-\frac{m_{s}^{3}\q2}{16\pi^{2}}\\
 \nonumber   &&+\frac{\G2m_{s}^{2}}{1024\pi^{6}}+\frac{\Gq m_{s}}{64\pi^{4}})s+(\frac{m_{s}^{2}\q2^{2}}{12\pi^{2}}-\frac{m_{s}^{2}\s2\q2}{4\pi^{2}}+\frac{m_{s}^{2}\s2^{2}}{48\pi^{2}}+\frac{\Gq m_{s}^{3}}{32\pi^{4}}+\frac{\Gs\q2}{24\pi^{2}}\\
 &&+\frac{\Gq\s2}{24\pi^{2}}-\frac{\G2\q2m_{s}}{384\pi^{4}}+\frac{\G2\s2m_{s}}{768\pi^{4}}),\\
\nonumber&&\\
\label{correusdd-TS} \rho^{a_{+} \rm OPE}_{T,S}&=&\rho^{a_{+}\rm
OPE}_{S,T}=-\frac{\G2s^{2}}{1024\pi^{6}}+\frac{3\G2m_{s}^{2}}{512\pi^{6}}s-(\frac{3\G2
m_{s}^{4}}{1024\pi^{6}}+\frac{\G2 m_{s}\s2}{128\pi^{4}}),
\end{eqnarray}
\begin{eqnarray}
\nonumber  \rho^{a_{+} \rm OPE}_{P,P}&=& \frac{s^{4}}{61440\pi^{6}}-\frac{m_{s}^{2}}{1536\pi^{6}}s^{3}+(\frac{m_{s}^{4}}{256\pi^{6}}+\frac{m_{s}(\s2+2\q2)}{192\pi^{4}}+\frac{\G2}{6144\pi^{6}})s^{2}-(\frac{\q2\s2}{12\pi^{2}}+\frac{m_{s}^{3}(\s2+2\q2)}{32\pi^{2}}\\
 \nonumber   &&+\frac{\G2m_{s}^{2}}{1024\pi^{6}}-\frac{\Gq m_{s}}{64\pi^{4}}-\frac{\Gs m_{q}}{64\pi^{4}})s+(\frac{3\s2\q2m_{s}^{2}}{4\pi^{2}}+\frac{\q2^{2}m_{s}^{2}}{12\pi^{2}}+\frac{m_{s}^{2}\s2^{2}}{48\pi^{2}}-\frac{\Gq m_{s}^{3}}{32\pi^{4}}\\
&&-\frac{\Gs\q2}{24\pi^{2}}-\frac{\Gq\s2}{24\pi^{2}}+\frac{\G2\q2m_{s}}{384\pi^{4}}+\frac{\G2\s2m_{s}}{768\pi^{4}},\\
\nonumber&&\\
\rho^{a_{+} \rm OPE}_{T,P}&=&\rho^{a_{+} \rm
OPE}_{P,T}=-\frac{\G2s^{2}}{1024\pi^{6}}+\frac{3\G2m_{s}^{2}}{512\pi^{6}}s-(\frac{3\G2
m_{s}^{4}}{1024\pi^{6}}+\frac{\G2m_{s}\s2}{128\pi^{4}}),\\
\nonumber&&\\
\nonumber  \rho^{a_{+} \rm OPE}_{A,A}&=& \frac{s^{4}}{7680\pi^{6}}-\frac{m_{s}^{2}}{192\pi^{6}}s^{3}+(\frac{m_{s}^{4}}{32\pi^{6}}+\frac{(\s2-\q2)m_{s}}{24\pi^{4}}+\frac{5\G2}{3072\pi^{6}})s^{2}-(\frac{5\G2m_{s}^{2}}{512\pi^{6}}+\frac{m_{s}^{3}(\s2-\q2)}{4\pi^{2}}\\
\nonumber && -\frac{\q2\s2}{3\pi^{2}}+\frac{5\Gq m_{s}}{16\pi^{4}})s+(\frac{2m_{s}^{2}\q2^{2}}{3\pi^{2}}-\frac{m_{s}^{2}\s2\q2}{\pi^{2}}+\frac{m_{s}^{2}\s2^{2}}{6\pi^{2}}+\frac{\Gq m_{s}^{3}}{8\pi^{4}}+\frac{\Gs\q2}{6\pi^{2}}\\
&&+\frac{\Gq\s2}{6\pi^{2}}-\frac{\G2\q2m_{s}}{192\pi^{4}}+\frac{5\G2m_{s}^{4}}{1024\pi^{6}}+\frac{5\G2\s2m_{s}}{384\pi^{4}}),\\
\nonumber&&\\
\rho^{a_{+} \rm OPE}_{A,S}&=&\rho^{a_{+} \rm OPE}_{S,A}=-\frac{3\G2m_{s}^{2}}{4096\pi^{6}}s+(\frac{\G2 m_{s}\s2}{256\pi^{4}}+\frac{\G2m_{s}\q2}{256\pi^{4}}),\\
\rho^{a_{+} \rm OPE}_{A,P}&=&\rho^{a_{+}\rm OPE}_{P,A}=0.
\end{eqnarray}
To convince the reader that our calculations are reliable, we make a
comparison with the results of other authors. For example,
\begin{eqnarray}
 \nonumber\rho^{\kappa \rm OPE}_{T,T}&=&\frac{s^4}{1280 \pi^6} -\frac{m_s^2 }
{64 \pi^6}s^3 +( \frac{11\G2} {768 \pi^6} +\frac{m_s\s2}{8 \pi^4})
s^2 - \frac{11 m_s^2 \G2}{256\pi^6}s + \frac{11
m_s\G2\s2}{192\pi^4}, \\
   \rho^{\kappa \rm OPE}_{S,S}&=&\frac{s^4} {61440 \pi^6}
-\frac{{m_s}^2 s^3} {3072 \pi^6} + ( \frac{\G2} {6144 \pi^6} -
\frac{{m_s}\q2} {192 \pi^4} + \frac{{m_s}\s2} {384 \pi^4}  ) s^2\\
\nonumber&& + ( - \frac{m_s^2 \G2}{2048\pi^6} - \frac{m_s \Gq
}{128\pi^4} + \frac{\q2^2}{24\pi^2} + \frac{\q2\s2 }{24\pi^2}) s
\\
 && - \frac{m_s^2 \q2^2}{12\pi^2} -
\frac{m_s\G2\q2}{768\pi^4} + \frac{m_s\G2\s2}{1536\pi^4} + \frac{\q2
\Gq}{24\pi^2} + \frac{\s2\Gq}{48\pi^2} + \frac{\q2\Gq}{48\pi^2}.
\end{eqnarray}
These are the expressions appearing in \cite{zhu2}.

\subsection{Instanton contribution to correlators of $\kappa$, $a_+$
and $f_0$} We obtain the intanton contributions to $\kappa$
correlators as follows,
\begin{eqnarray}\label{}
    \Pi_{TT}^{\kappa(\rm inst)}&=&(\frac{76n_{eff}r_{c}^{4}\q2\s2}{3\pi^{4}m_{q}^{\ast2}}+\frac{144n_{eff}r_{c}^{4}\q2^{2}}{3\pi^{4}m_{q}^{\ast}m_{s}^{\ast}})f_{0}(Q),\\
  \Pi_{SS}^{\kappa(\rm inst)}&=&(\frac{16n_{eff}r_{c}^{4}}{\pi^{8}m_{q}^{\ast}m_{s}^{\ast}}+\frac{16n_{eff}r_{c}^{4}}{\pi^{8}m_{q}^{\ast2}})f_{6}(Q)+(\frac{11n_{eff}r_{c}^{4}\q2^{2}}{18\pi^{4}m_{q}^{\ast}m_{s}^{\ast}}+\frac{19n_{eff}r_{c}^{4}\q2\s2}{36\pi^{4}m_{q}^{\ast2}})f_{0}(Q),\\
\nonumber \Pi_{PP}^{\kappa(\rm inst)}&=&-(\frac{16n_{eff}r_{c}^{4}}{\pi^{8}m_{q}^{\ast}m_{s}^{\ast}}+\frac{16n_{eff}r_{c}^{4}}{\pi^{8}m_{q}^{\ast2}})f_{6}(Q)+(\frac{19n_{eff}r_{c}^{4}\q2\s2}{36\pi^{4}m_{q}^{\ast2}}+\frac{11n_{eff}r_{c}^{4}\q2^{2}}{18\pi^{4}m_{q}^{\ast}m_{s}^{\ast}})f_{0}(Q),\\
\Pi_{TS}^{\kappa(\rm inst)}&=&\Pi_{ST}^{\kappa(\rm inst)}=\frac{n_{eff}r_{c}^{4}\q2\s2}{\pi^{4}m_{q}^{\ast2}}f_{0}(Q),\\
   \Pi_{TP}^{\kappa(\rm inst)}&=&\Pi_{PT}^{\kappa(\rm
   inst)}=\frac{n_{eff}r_{c}^{4}\q2\s2}{\pi^{4}m_{q}^{\ast2}}f_{0}(Q),\\
  \Pi_{AA}^{\kappa(\rm
  inst)}&=&(\frac{24n_{eff}r_{c}^{4}}{\pi^{8}m_{q}^{\ast}m_{s}^{\ast}}+\frac{24n_{eff}r_{c}^{4}}{\pi^{8}m_{q}^{\ast2}})f_{6}(Q)+(\frac{37n_{eff}r_{c}^{4}\q2^{2}}{6\pi^{4}m_{q}^{\ast}m_{s}^{\ast}}+\frac{34n_{eff}r_{c}^{4}\q2\s2}{9\pi^{4}m_{q}^{\ast2}})f_{0}(Q),\\
    \Pi_{AS}^{\kappa(\rm inst)}&=&\Pi_{SA}^{\kappa(\rm inst)}=-(\frac{20n_{eff}r_{c}^{4}}{\pi^{8}m_{q}^{\ast}m_{s}^{\ast}}+\frac{10n_{eff}r_{c}^{4}}{\pi^{8}m_{q}^{\ast2}})f_{6}(Q),\\
    \Pi_{AP}^{\kappa(\rm inst)}&=&\Pi_{PA}^{\kappa(\rm
    inst)}=-(\frac{20n_{eff}r_{c}^{4}}{\pi^{8}m_{q}^{\ast}m_{s}^{\ast}}+\frac{10n_{eff}r_{c}^{4}}{\pi^{8}m_{q}^{\ast2}})f_{6}(Q).
\end{eqnarray}

The instanton contributions to $a_{+}$ are,
\begin{eqnarray}\label{}
    \Pi_{TT}^{a_{+}(\rm inst)}&=&(\frac{152n_{eff}r_{c}^{4}\q2\s2}{3\pi^{4}m_{q}^{\ast}m_{s}^{\ast}}+\frac{68n_{eff}r_{c}^{4}\s2^{2}}{3\pi^{4}m_{q}^{\ast2}})f_{0}(Q),\\
  \Pi_{SS}^{a_{+}(\rm inst)}&=&
  \frac{32n_{eff}r_{c}^{4}}{\pi^{8}m_{q}^{\ast}m_{s}^{\ast}}f_{6}(Q)+(\frac{19n_{eff}r_{c}^{4}\q2\s2}{18\pi^{4}m_{q}^{\ast}m_{s}^{\ast}}+\frac{n_{eff}r_{c}^{4}\s2^{2}}{12\pi^{4}m_{q}^{\ast2}})f_{0}(Q),\\
    \Pi_{PP}^{a_{+}(\rm inst)}&=&-\frac{32n_{eff}r_{c}^{4}}{\pi^{8}m_{q}^{\ast}m_{s}^{\ast}}f_{6}(Q)+(\frac{19n_{eff}r_{c}^{4}\q2\s2}{18\pi^{4}m_{q}^{\ast}m_{s}^{\ast}}+\frac{n_{eff}r_{c}^{4}\s2^{2}}{12\pi^{4}m_{q}^{\ast2}})f_{0}(Q),\\
    \Pi_{TS}^{a_{+}(\rm inst)}&=&\Pi_{ST}^{a_{+}(\rm inst)}=(\frac{2n_{eff}r_{c}^{4}\q2\s2}{\pi^{4}m_{q}^{\ast}m_{s}^{\ast}}-\frac{n_{eff}r_{c}^{4}\s2^{2}}{\pi^{4}m_{q}^{\ast2}})f_{0}(Q),\\
   \Pi_{TP}^{a_{+}(\rm inst)}&=&\Pi_{PT}^{a_{+}(\rm inst)}=(\frac{2n_{eff}r_{c}^{4}\q2\s2}{\pi^{4}m_{q}^{\ast}m_{s}^{\ast}}-\frac{n_{eff}r_{c}^{4}\s2^{2}}{\pi^{4}m_{q}^{\ast2}})f_{0}(Q),\\
    \Pi_{AA}^{a_{+}(\rm inst)}&=&\frac{48n_{eff}r_{c}^{4}}{\pi^{8}m_{q}^{\ast}m_{s}^{\ast}}f_{6}(Q)+(\frac{68n_{eff}r_{c}^{4}\q2\s2}{9\pi^{4}m_{q}^{\ast}m_{s}^{\ast}}+\frac{43n_{eff}r_{c}^{4}\s2^{2}}{18\pi^{4}m_{q}^{\ast2}})f_{0}(Q),\\
    \Pi_{AS}^{a_{+}(\rm inst)}&=&\Pi_{SA}^{a+(\rm inst)}=-(\frac{20n_{eff}r_{c}^{4}}{\pi^{8}m_{q}^{\ast}m_{s}^{\ast}}+\frac{10n_{eff}r_{c}^{4}}{\pi^{8}m_{q}^{\ast2}})f_{6}(Q),\\
    \Pi_{AP}^{a_{+}(\rm inst)}&=&\Pi_{PA}^{a_{+}(\rm inst)}=-(\frac{20n_{eff}r_{c}^{4}}{\pi^{8}m_{q}^{\ast}m_{s}^{\ast}}+\frac{10n_{eff}r_{c}^{4}}{\pi^{8}m_{q}^{\ast2}})f_{6}(Q).
\end{eqnarray}

The instanton contributions to $f_{0}$ are,
\begin{eqnarray}\label{}
    \Pi_{TT}^{f_0(\rm inst)}&=&(\frac{152n_{eff}r_{c}^{4}\q2\s2}{3\pi^{4}m_{q}^{\ast}m_{s}^{\ast}}-\frac{68n_{eff}r_{c}^{4}\s2^{2}}{3\pi^{4}m_{q}^{\ast2}})f_{0}(Q),\\
  \Pi_{SS}^{f_0(\rm inst)}&=&
  \frac{32n_{eff}r_{c}^{4}}{\pi^{8}m_{q}^{\ast}m_{s}^{\ast}}f_{6}(Q)+(\frac{19n_{eff}r_{c}^{4}\q2\s2}{18\pi^{4}m_{q}^{\ast}m_{s}^{\ast}}-\frac{n_{eff}r_{c}^{4}\s2^{2}}{12\pi^{4}m_{q}^{\ast2}})f_{0}(Q),\\
    \Pi_{PP}^{f_0(\rm inst)}&=&-\frac{32n_{eff}r_{c}^{4}}{\pi^{8}m_{q}^{\ast}m_{s}^{\ast}}f_{6}(Q)+(\frac{19n_{eff}r_{c}^{4}\q2\s2}{18\pi^{4}m_{q}^{\ast}m_{s}^{\ast}}-\frac{n_{eff}r_{c}^{4}\s2^{2}}{12\pi^{4}m_{q}^{\ast2}})f_{0}(Q),\\
    \Pi_{TS}^{f_0(\rm inst)}&=&\Pi_{ST}^{f_0(\rm inst)}=(\frac{2n_{eff}r_{c}^{4}\q2\s2}{\pi^{4}m_{q}^{\ast}m_{s}^{\ast}}+\frac{n_{eff}r_{c}^{4}\s2^{2}}{\pi^{4}m_{q}^{\ast2}})f_{0}(Q),\\
   \Pi_{TP}^{f_0(\rm inst)}&=&\Pi_{PT}^{f_0(\rm
   inst)}=(\frac{2n_{eff}r_{c}^{4}\q2\s2}{\pi^{4}m_{q}^{\ast}m_{s}^{\ast}}+\frac{n_{eff}r_{c}^{4}\s2^{2}}{\pi^{4}m_{q}^{\ast2}})f_{0}(Q),\\
    \Pi_{AA}^{f_0(\rm inst)}&=&\frac{48n_{eff}r_{c}^{4}}{\pi^{8}m_{q}^{\ast}m_{s}^{\ast}}f_{6}(Q)+(\frac{68n_{eff}r_{c}^{4}\q2\s2}{9\pi^{4}m_{q}^{\ast}m_{s}^{\ast}}-\frac{43n_{eff}r_{c}^{4}\s2^{2}}{18\pi^{4}m_{q}^{\ast2}})f_{0}(Q),\\
    \Pi_{AS}^{f_0(\rm inst)}&=&\Pi_{SA}^{f_0(\rm inst)}=-(\frac{20n_{eff}r_{c}^{4}}{\pi^{8}m_{q}^{\ast}m_{s}^{\ast}}-\frac{10n_{eff}r_{c}^{4}}{\pi^{8}m_{q}^{\ast2}})f_{6}(Q),\\
    \Pi_{AP}^{f_0(\rm inst)}&=&\Pi_{PA}^{f_0(\rm inst)}=-(\frac{20n_{eff}r_{c}^{4}}{\pi^{8}m_{q}^{\ast}m_{s}^{\ast}}-\frac{10n_{eff}r_{c}^{4}}{\pi^{8}m_{q}^{\ast2}})f_{6}(Q).
\end{eqnarray}
To check our results, we take the $SU(3)_f$ limit, which is
$m^\ast_q=m^\ast_s$ and $\q2=\s2$. In this limit, the instanton
contributions to $\kappa$ and $a_+$ are equal to each other. This is
because that they all belong to the octet representation of
$SU(3)_f$. But this check is not suitable for $f_0$, since it comes
from ideally mixing of the flavor singlet state with the isospin
$I$=0 component of flavor octet state .
\begin{table}[hbt]
\caption{Fitted masses and residues in single resonance
approximation for $\kappa$}
\begin{center}
\begin{tabular}{c|c|c}
 \hline  $s_{0}({\rm GeV}^{2})$ & $M_{\kappa}$(GeV) & ${ f}_{\kappa}  (10^{-2}\rm GeV)$ \\
 \hline   1 & 0.72 & 1.04  \\
 \hline   1.5  & 0.73 &1.00 \\
  \hline
\end{tabular}
\end{center}
\end{table}
\begin{table}[hbt]
\caption{Fitted masses and residues in single resonance
approximation for $a_{+}$}
\begin{center}
\begin{tabular}{c|c|c}
 \hline  $s_{0}({\rm GeV}^{2})$ & $M_{a_{+}}$(GeV) & ${ f}_{a_{+}}  (10^{-2}\rm GeV)$ \\
 \hline   1 & 0.73 & 0.89  \\
 \hline   1.5  & 0.73 &0.90 \\
  \hline
\end{tabular}
\end{center}
\caption{Fitted masses and residues in single resonance
approximation for $f_{0}$}
\begin{center}
\vspace{0ex}
\begin{tabular}{c|c|c}
 \hline  $s_{0}({\rm GeV}^{2})$ & $M_{f_{0}}$(GeV) & ${ f}_{f_{0}}  (10^{-2}\rm GeV)$ \\
 \hline   1 & 0.72 & 0.90  \\
 \hline   1.5  & 0.72 &0.94 \\
  \hline
\end{tabular}
\end{center}
\end{table}
\end{appendix}

\end{document}